\documentclass{jfm}
\usepackage{graphicx}
\usepackage{epstopdf, epsfig}
\usepackage[utf8]{inputenc}
\usepackage{tikz}
        \usetikzlibrary{calc}
        \usepackage{pgfplots}
\usepackage{caption}
\usepackage{subcaption}
\usepackage{adjustbox}
\usepackage{amsmath}

\newcommand{\Y}{\Gamma}
\newcommand{\grad}{\textrm{grad}_M}
\newcommand{\gradc}{\boldsymbol{\nabla}_c}
\newcommand{\gradcc}{\nabla_c}
\renewcommand{\div}{\textrm{div}_M}
\newcommand{\divb}{\textbf{div}_M}
\newcommand{\curl}{\textrm{curl}_M}
\newcommand{\laplace}{\boldsymbol{\Delta}_B}
\newcommand{\qlaplace}{\boldsymbol{\Delta}_Q}

\newcommand{\hatgradc}{\hat{\boldsymbol{\nabla}}_c}
\newcommand{\hatdiv}{\widehat{\textrm{div}}_M}
\newcommand{\hatlaplace}{\hat{\boldsymbol{\Delta}}_B}

\renewcommand{\u}{\boldsymbol{u}}
\renewcommand{\v}{\boldsymbol{v}}
\newcommand{\w}{\boldsymbol{w}}
\newcommand{\x}{\boldsymbol{x}}

\renewcommand{\a}{\boldsymbol{a}}
\newcommand{\pim}{\boldsymbol{\pi}_M}
\newcommand{\pimh}{\boldsymbol{\pi}_{M_h}}
\newcommand{\nub}{\boldsymbol{\nu}}
\newcommand{\R}{\mathbb{R}}

\newcommand{\integrate}[3]{\int_{#1}^{} #2 \, \mathrm{d} #3}
\newcommand{\intmh}[1]{\integrate{M_h}{#1}{A_h} }
\newcommand{\bracket}[1]{\left(#1\right)}
\newcommand{\Q}{\boldsymbol{\nabla}_Q}
\newcommand{\uk}{u^*_k}
\newcommand{\vk}{v^*_k}
\newcommand{\wk}{w^*_k}
\newcommand{\unk}{u^n_k}
\newcommand{\vnk}{v^n_k}
\newcommand{\wnk}{w^n_k}

\pgfplotsset{compat=1.16}

\shorttitle{Active flows on curved surfaces}
\shortauthor{M. Rank and A. Voigt}

\title{Active Flows on Curved Surfaces}

\author{M. Rank\aff{1} \and A. Voigt\aff{1,2,3}\corresp{\email{axel.voigt@tu-dresden.de}}}

\affiliation{\aff{1}Institut für Wissenschaftliches Rechnen, TU Dresden, 01062 Dresden, Germany \aff{2}Center for Systems Biology Dresden (CSBD), Pfotenhauerstr. 108, 01307 Dresden, Germany \aff{3} Cluster of Excellence - Physics of Life, TU Dresden, 01062 Dresden, Germany}

\begin{document}

\maketitle

\begin{abstract}
We consider a numerical approach for a covariant generalised Navier-Stokes equation on general surfaces and study the influence of varying Gaussian curvature on anomalous vortex-network active turbulence. This regime is characterised by self-assembly of finite-size vortices into linked chains of anti-ferromagnet order, which percolate through the entire surface. The simulation results reveal an alignment of these chains with minimal curvature lines of the surface and indicate a dependency of this turbulence regime on the sign and the gradient in local Gaussian curvature. While these results remain qualitative and their explanations are still incomplete, several of the observed phenomena are in qualitative agreement with experiments on active nematic liquid crystals on toroidal surfaces and contribute to an understanding of the delicate interplay between geometrical properties of the surface and characteristics of the flow field, which has the potential to control active flows on surfaces via gradients in the spatial curvature of the surface.
\end{abstract}


\section{Introduction}\label{sec:introduction}

To model fluids on curved surfaces is a problem which dates back to \citet{Scriven_CES_1960}, who derived a covariant Navier-Stokes (NS) equation and established the coupling between spatial curvature and fluid flow. The influence of geometric properties on both, equilibrium configurations and the dynamics far from it, has consequences in a huge variety of problems, ranging from planetary flows \citep{Delplace_Science_2017} to active nematic films on vesicles \citep{Keber_Science_2014}. 
Further examples, where the spatial curvature influences fluid flow, are found in developmental biology, e.g. tissue morphogenesis \citep{Heisenberg_Cell_2013}, cell division \citep{Mayer_Nature_2010} and biochemical signal propagation \citep{Tan_NP_2020}, biofilm formation \citep{Chang_NJP_2015} and bacterial colonization \citep{Sipos_PRL_2015}. All these examples offer the possibility to influence or even control fluid flow by the spatial curvature of the surface. 

The resulting huge interest in surface flows is in contrast to a still missing coherent theoretical understanding of the interplay with geometric properties. Analytical results in oversimplified situations are of limited value in this context and also numerical approaches were until recently restricted to special cases. Most of them are based on a vorticity-stream function formulation \citep{Nitschke_JFM_2012,Reuther_MMS_2015,Gross_JCP_2018,Reusken_IMAJNA_2020} and have been applied to surface Stokes or NS equations. More recently they are also considered simulating active flows on surfaces \citep{Pearce_PRL_2019,Torres-Sanchez_JFM_2019,Supekar_JFM_2020}. The advantage of the vorticity-stream function formulation is the avoidance of tangential vector fields. This allows to explore well-established numerical methods for scalar fields on surfaces. However, in contrast to the Euclidian setting, the Helmholtz-Hodge decomposition - see \citet{Bhatia_IEEE_2013} for a review - which provides the mathematical basis for the vorticity-stream function formulation, not only splits the tangential velocity field into curl-free and divergence-free components, but might also contain non-trivial harmonic vector fields - vector fields which are curl- and divergence-free. Their richness depends on the topology of the surface \citep{Jost_book_1991}. As these vector fields cannot be described by the vorticity-stream function formulation, the approach is only applicable for surfaces, where harmonic vector fields are trivial, which are only simply-connected surfaces  \citep{Nitschke_PAMM_2021}. \citet{Reuther_PF_2018} and \citet{Fries_IJNMF_2018} introduce a numerical approach for a surface NS equation in its velocity-pressure formulation, which is applicable on general surfaces. The underlying idea was independently introduced in \citet{Nestler_JNS_2018}, \citet{Jankuhn_IFB_2018} and \citet{Hansbo_IMAJNA_2020} and is described in detail in \citet{Nestler_JCP_2019}. It relies on an extension to an embedding thin film. This allows to express the covariant derivatives in terms of partial derivatives along the Euclidian basis. The surface partial differential equation can thus be solved by considering each component by established methods for scalar quantities and enforcing the normal contributions to be zero, either by penalisation or by Lagrange multipliers. 
Also this approach has been applied to active flows in the context of surface active polar gels \citep{Nitschke_PRF_2019}. Other numerical approaches for the surface Stokes or NS equation, directly acting on the tangential velocity and pressure fields, are proposed in \citet{Nitschke_book_2017}, \citet{Torres-Sanchez_JCP_2020}, \citet{Sahu_JCP_2020} and \citet{Lederer_arXiv_2019}.

We will here consider a modelling approach for active flows and extend studies for a generalised Navier-Stokes (GNS) equation on a sphere \citep{Mickelin_PRL_2018} to toroidal surfaces. This GNS equation describes internally driven flows through higher-order hyperviscosity-like terms in the stress tensor. In flat space such models have been phenomenologically proposed for active fluids \citep{Wensink_PNAS_2012} with additional Toner-Tu like terms \citep{Toner_PRE_1998} and later on have been justified by microscopic theories \citep{Heidenreich_PRE_2016}. The proposed version in \citet{Slomka_EPJST_2015,Slomka_PRF_2017} focuses on the solvent velocity field. Under the assumption of dense suspensions this allows to neglect the local driving terms and additional active stresses and describe the active flow solely by a generalised stress tensor, which comprises passive contributions from the intrinsic solvent fluid viscosity and active contributions representing the stresses exerted by the active components on the fluid. The version in \citet{Slomka_EPJST_2015,Slomka_PRF_2017} can also be derived from classical hydrodynamic models with active stresses \citep{Linkmannetal_PRE_2020}. The model serves as a minimal model for active fluids and had been successful in reproducing active turbulence flow patterns of swimming bacteria, ATP-driven microtubules and artificial microswimmers, see, e.g. \citep{Bratanovetal_PNAS_2015,Jamesetal_PRF_2018} and had also been studied previously for seismic wave propagation \citep{Beresnev_PD_1993,Tribelsky_PRL_1996}. Investigations on surfaces are so far restricted to spherical surfaces \citep{Mickelin_PRL_2018,Supekar_JFM_2020}. The considered numerical approach uses a pseudo-spectral method with a basis of spherical harmonics. Besides the restriction of this approach to a sphere, also the constant Gaussian curvature in this case implies that only the size of the sphere matters. Non of these approaches is applicable to general surfaces.

Results on the influence of varying Gaussian curvature on active flows are rare. Besides a few microscopic models for active nematics, which do not account for hydrodynamic effects \citep{Alaimo_SR_2017,Apaza_SM_2018,Ellis_NP_2018}, continuum models, with numerics based on the vorticity-stream function formulation \citep{Torres-Sanchez_JFM_2019,Pearce_PRL_2019} and the generally applicable approach for polar liquid crystals \citep{Nitschke_PRF_2019} as well as first experimental work on active nematic liquid
crystals comprised of microtubules and kinesin, which are constrained to lie on a toroidal surface \citep{Ellis_NP_2018}, no results are available. The last is especially interesting as it has regions with both positive and negative Gaussian curvature. From equilibrium considerations it is expected that topological defects in the nematic liquid crystal, here of charge $\pm\frac{1}{2}$, are attracted by regions of like-sign Gaussian curvature. A phenomenon which has been explored in detail for disclinations in positional order \citep{Bowick_PRE_2004,Giomi_EPJE_2008} and is expected to carry over to orientational order \citep{Turner_RMP_2010,Jesenek_SM_2015}. The experiments in \citet{Ellis_NP_2018} suggest that active flows on surfaces can be guided and controlled via gradients in the spatial curvature of the surface. They show that pairs of defects unbind and segregate in regions of opposite Gaussian curvature. At least qualitatively these results have been reproduced in \citet{Pearce_PRL_2019}, showing a linear dependency of defect creation and annihilation rates on Gaussian curvature. As these rates are directly linked to active turbulence \citep{Giomi_PRX_2015}, a connection between spatial curvature and active turbulence can be expected. \citet{Pearce_PRL_2019} consider an active nematodynamics model which is based on a simplified surface Landau-de Gennes energy \citep{Kralj_SM_2011}. For more detailed surface Landau-de Gennes models which also take extrinsic curvature contributions into account, see \citet{Golovary_JNS_2017}, \citet{Nitschke_PRSA_2018} and \citet{Nestler_SM_2020}. As already pointed out, the numerical treatment in \citet{Pearce_PRL_2019} is based on a vorticity-stream function formulations, which is inappropriate for toroidal surfaces. Appropriate numerical studies for active nemotodynamics on general surfaces are still under development. However, several open questions, e.g., the influence of spatial curvature on active turbulence can already be studies using an appropriately coarse-grained mesoscopic GNS equation. 

Here we focus on one aspect of the active flow. We only consider the influence of spatial varying curvature on anomalous vortex-network active turbulence. The simulations in \citet{Mickelin_PRL_2018} on a sphere reveal a global curvature-induced transition from a quasi-stationary burst phase to an anomalous vortex-network turbulent phase and a classical 2D Kolmogorov turbulent phase. The new type of anomalous turbulence is characterised by the self-assembly of finite-size vortices into linked chains of anti-ferromagnetic order, which percolate through the entire fluid domain. The coherent motion of this vortex chain network provides an upward energy transfer and thus an alternative to the conventional energy cascade in classical 2D hydrodynamic turbulence. We will answer the question if this mechanism is altered by gradients in curvature by considering the GNS equation on different tori within a parameter setting which leads to anomalous vortex-network turbulence on a sphere.

The paper is organised as follows: In Section \ref{sec:mathematical_model} we introduce a covariant formulation of the GNS equation which is applicable to general curved surfaces. In Section \ref{sec:numerical_approach} we describe the numerical approach, including basic validations for NS and GNS equations. The discussion on the influence of curvature on anomalous vertex-network turbulence is done in Section \ref{sec:results} and conclusions are drawn in Section \ref{sec:conclusion}.

\section{Mathematical model}\label{sec:mathematical_model}
In \citet{Mickelin_PRL_2018} the surface generalised Navier-Stokes (GNS) equation 
\begin{eqnarray}
    \p_t \u(\x,t)+\boldsymbol{\nabla}_{\u} \u(\x,t) &=& -\grad p(\x,t) + \divb\boldsymbol{T}, \label{eq:general stokes}\\
    \div \u(\x,t) &=& 0 \label{eq:general div}
\end{eqnarray}
with initial condition $\u(\x,0)=\u_{0}(\x)\in T_x M$ was proposed on $M\times(0,\infty)$ with $M$ a sphere. The tangential fluid velocity at point $\x\in M$ and time $t\in(0,\infty)$ is denoted by $\u(\x,t)\in TM$ and the surface pressure by $p(\x,t)\in\R$. \citet{Mickelin_PRL_2018} considers the surface tension $-p(\x,t)$. $\partial_t$ is the time derivative, $\boldsymbol{\nabla}_{\u}$ the covariant directional derivative, $\grad$ the surface gradient, and $\div$ and $\divb$ the surface divergence, for vector- and tensor-fields, respectively. The stress tensor $\boldsymbol{T}$ contains passive and active contributions and reads 
\begin{eqnarray}
\boldsymbol{T} &=& f\bracket{\boldsymbol{\Delta}_M}\bracket{\grad \u + \bracket{\grad \u}^T}, \label{eq:stress}\\
f\bracket{\boldsymbol{\Delta}_M} &=& \Y_0-\Y_2\boldsymbol{\Delta}_M+\Y_4\boldsymbol{\Delta}_M^2, \label{eq:f}
\end{eqnarray}
where $\boldsymbol{\Delta}_M^2=\boldsymbol{\Delta}_M\boldsymbol{\Delta}_M$ with a surface Laplacian $\boldsymbol{\Delta}_M$ and real parameters $\Y_0,\Y_2,\Y_4\in\R$. The constants $\Y_0$ and $\Gamma_4$ are assumed to be positive to ensure asymptotic stability, whereas $\Y_2$ may have either sign. For $\Y_2 < 0$ nontrivial steady-state flow structures may emerge. Viewing $f(\boldsymbol{\Delta}_M)$ as an effective viscosity, sufficiently negative $\Y_2$ may turn this quantity negative and thus providing a source of energy which makes the model effectively active. We can think about $\boldsymbol{\Delta}_M$ as a wildcard for any surface Laplacian. \citet{Mickelin_PRL_2018} have been using the Bochner Laplace operator $\boldsymbol{\Delta}_M=\laplace$. Other choices are the Laplace-deRham operator $\boldsymbol{\Delta}_{dR}$ or the Q-Laplacian $\qlaplace$. They are related by $\qlaplace \u = \laplace \u + \kappa \u = \boldsymbol{\Delta}_{dR} \u + 2 \kappa \u$, with $\kappa$ the Gaussian curvature \citep{Abraham2012}. In flat space they are all identical and also for the sphere, where the Gaussian curvature $\kappa$ is constant, evaluating $\divb\boldsymbol{T}$ is not an issue for any of the operators. However, evaluating $\divb\boldsymbol{T}$ on more general surfaces, like the torus, turns out to be most convenient using the Q-Laplacian. This leads to the following computation:
\begin{eqnarray}
    \divb\boldsymbol{T} &=& \divb f\bracket{\qlaplace}(\grad \u + \bracket{\grad \u}^T) \nonumber \\
    &=& \divb f\bracket{\qlaplace}\cdot 2\boldsymbol{\nabla}_Q \u \nonumber \\
    &=& 2\Y_0\divb\Q \u - 2\Y_2\divb\qlaplace\Q \u + 2\Y_4\divb\qlaplace^2\boldsymbol{\nabla}_Q \u \nonumber \\
    &=& \Y_0\cdot 2\divb\Q \u - \Y_2\cdot 2\divb\Q\qlaplace \u + \Y_4\cdot 2\divb\boldsymbol{\nabla}_Q\qlaplace^2 \u \nonumber \\
    &=& \Y_0 \qlaplace \u - \Y_2 \qlaplace^2 \u + \Y_4 \qlaplace^3 \u, \label{eq:divergence surface stress tensor}
\end{eqnarray}
where $\qlaplace^3=\qlaplace\qlaplace\qlaplace$. Introducing the auxiliary quantities $\v=\qlaplace \u$ and $\w=\qlaplace \v = \qlaplace^2 \u$, using eq. \eqref{eq:divergence surface stress tensor} and the Bochner Laplacian $\laplace$, eqs. \eqref{eq:general stokes} and \eqref{eq:general div} can be rewritten as a system of second order partial differential equations
\begin{eqnarray}
    \!\!\!\!\!\!\!\!\!\!\!\!\!\!\partial_t \u+\boldsymbol{\nabla}_{\u} \u &=&-\grad p + \Y_0\bracket{\laplace \u + \kappa \u} - \Y_2\bracket{\laplace \v + \kappa \v} +\Y_4\bracket{\laplace \w + \kappa \w}, \label{eq:active ns}\\
    \v &=&\laplace \u + \kappa \u, \label{eq:v}\\
    \w &=&\laplace \v + \kappa \v, \label{eq:w}\\
    \div \u &=&0, \label{eq:div}
\end{eqnarray}
which we will consider as a surface GNS equation on $M$, with $M$ now a compact smooth 2-manifold without boundary. In the case of constant Gaussian curvature $\kappa$ the model coincides with the model presented in \citet{Mickelin_PRL_2018} with the particular choice of $\Y_0=\overline{\Y_0}-2\overline{\Y_2}\kappa+4\overline{\Y_4}\kappa^2$, $\Y_2=\overline{\Y_2}-4\overline{\Y_4}\kappa$ and $\Y_4=\overline{\Y_4}$. Thereby $\overline{\Y_0}$, $\overline{\Y_2}$ and $\overline{\Y_4}$ denote the according parameters introduced in \citet{Mickelin_PRL_2018}. For $\Y_2 = \Y_4 = 0$ we obtain as a special case the surface NS equation \citep{Scriven_CES_1960}. This limit, which also follows as a thin-film limit of the 3D NS equation with Navier boundary conditions \citep{Miura_QAM_2018,Nitschke_PRF_2019,Reuther_JFM_2020} justifies the choice of the surface Laplacian in the derivation. We further expect also to obtain eqs. \eqref{eq:active ns}-\eqref{eq:div} as a thin-film limit of the 3D GNS with Navier boundary conditions. However, this analysis has not been done. 

It is not obvious to see why eqs. \eqref{eq:active ns} - \eqref{eq:div} provide an effective model for internally driven flows. The active component is hidden in the effective viscosity $f\bracket{\qlaplace}$, which accounts for the intrinsic solvent fluid viscosity and contributions representing the stresses exerted by the active components on the fluid. For $\Gamma_2 < 0$ an instability can occur and the interplay between this instability and the nonlinerity of the surface NS equations drives the spatiotemporal dynamics and leads in certain parameter regimes to the formation of mesoscale vortices. Due to the additional coupling with geometric quantities we expect the vortices and the parameter regime to be effected by the Gaussian curvature of the surface. While for constant Gaussian curvature $\kappa$ exact stationary solutions can be constructed, which show some aspects of this influence \citep[see][]{Mickelin_PRL_2018}, for the general case we have to rely on numerical approximations. 

\section{Numerical approach}\label{sec:numerical_approach}

Following the general approach \citep{Nestler_JCP_2019} we first reformulate eqs. \eqref{eq:active ns} - \eqref{eq:div} to a system which fulfils $\u(\x,t)\in TM$ only approximately. This system can then be numerically solved using a time discretisation based on a Chorin projection method and a space discretisation by a regular surface triangulation and a scalar-valued surface finite element method \citep{Dziuk_AN_2013} applied to each component of the extended velocity field, each component of the extended auxiliary variables and the pressure field. The approach extends the surface finite element discretisation for the surface NS equation \citep{Reuther_PF_2018} to the surface GNS equation. 

\subsection{Reformulation}

Instead of the tangential fields $\u(\x,t), \v(\x,t), \w(\x,t)\in TM$ in the surface GNS equation \eqref{eq:active ns} - \eqref{eq:div} we will consider $\R^3$-valued vector fields $\hat{\u}(\x,t), \hat{\v}(\x,t), \hat{\w}(\x,t)\in\R^3$ which are only weakly tangential to $TM$. We thereby approximate the surface GNS equation following the general method proposed by \citet{Nestler_JCP_2019}. We consider a neighbourhood $U_\delta$ of $M$ with coordinate projection $\boldsymbol{\pi}:U_\delta\to M$ of $\tilde{\x}=\a(\tilde{\x}) +d(\tilde{\x})\nub\left(\a(\tilde{\x})\right)\in U_\delta$, with $d$ the signed distance function and $\nub$ the surface normal, given by $\tilde{\x}\mapsto\x=\a(\tilde{\x})\in M$. Depending on the curvature of the surface $M$, this coordinate projection is injective for $\delta>0$ small enough. The velocity field, the auxiliary variables and the pressure are smoothly extended to $U_\delta$ such that $\tilde{\u}(\tilde{\x})=\u\bracket{\x}$, $\tilde{\v}(\tilde{\x})=\v\bracket{\x}$, $\tilde{\w}(\tilde{\x})=\w\bracket{\x}$ and $\tilde{p}(\tilde{\x})=p\bracket{\x}$ are obtained for all $\tilde{\x}\in U_\delta$. We extend the tangential differential operators by
\begin{eqnarray*}
    \hatgradc\hat{\u} = \pim\cdot\bracket{\boldsymbol{\nabla}\hat{\u} -\boldsymbol{\nabla}\hat{\u}\cdot\nub\nub}, \quad
    \hatdiv\hat{\u} = \textrm{div}\bracket{\hat{\u}} - \nub\cdot\bracket{\boldsymbol{\nabla}\hat{\u}\cdot\nub}, \quad
    \hatlaplace\hat{\u} = \widehat{\textbf{div}}_M \hatgradc\hat{\u} \label{eq:extended laplace},
\end{eqnarray*}
where $\boldsymbol{\nabla}$, $\textrm{div}$ and $\textbf{div}$ denote the common vector gradient and divergence operators in $\R^3$. For the general form for $\widehat{\textbf{div}}_M$ we refer to \citet{Nestler_JCP_2019}(Eq. E.2). The pointwise normal projection is given by $\pim=\boldsymbol{I}-\nub\nub^T$, where $\boldsymbol{I}$ denotes the identity matrix.
Adding $\alpha(\nub\cdot\hat{\u})\nub$, $\alpha(\nub\cdot\hat{\v})\nub$ and $\alpha(\nub\cdot\hat{\w})\nub$ with penalty parameter $\alpha\in\R$ big enough penalises the normal components of the vectors $\hat{\u}$, $\hat{\v}$ and $\hat{\w}$, see \citet[]{Nestler_JCP_2019}. This motivates the choice of the operators above, as we assume the velocity $\hat{\u}$ to be approximately tangential to the surface. In \citet{Nestler_JCP_2019} it was shown that $\hatlaplace\hat{\u}\approx\laplace \u$, when using an appropriate penalty term. For convergence studies and possible dependencies of $\alpha$ on mesh size, we refer to \citet{Hansbo_IMAJNA_2020}. \citet{Reuther_PF_2018} motivate the replacement of the surface divergence $\div \u$ by $\hatdiv \hat{\u}$. This leads to the following approximation of the surface GNS equation in Cartesian coordinates:
\begin{eqnarray}
    \p_t \hat{\u}+ \hat{\u} \cdot \hatgradc \hat{\u} &=&-\hatgradc p + \Y_0 \hat{\v} - \Y_2 \hat{\w} +\Y_4(\hatlaplace \hat{\w} + \kappa \hat{\w}) -\alpha(\nub\cdot\hat{\u})\nub, \label{eq:penalty active ns}\\
    \hat{\v} &=&\hatlaplace \hat{\u} + \kappa \hat{\u} -\alpha(\nub\cdot\hat{\v})\nub, \label{eq:penalty v}\\
    \hat{\w} &=&\hatlaplace \hat{\v} + \kappa \hat{\v}-\alpha(\nub\cdot\hat{\w})\nub, \label{eq:penalty w}\\
    \hatdiv \hat{\u} &=&0. \label{eq:penalty div}
\end{eqnarray}
The above model formulation coincides with the initial one just in the case of $\nub\cdot\hat{\u}=\nub\cdot\hat{\v}=\nub\cdot\hat{\w}=0$ and ensures only a weak form of tangency of the according solution $\hat{\u}$ \citep{Reuther_PF_2018,Nestler_JCP_2019}. Hereafter, we assume the vector fields $\hat{\u}$, $\hat{\v},$ and $\hat{\w}$ to be tangential to the surface, which is legitimate for appropriate $\alpha$ \citep{Reuther_PF_2018,Hansbo_IMAJNA_2020}. Operators as well as functions will remain to have the same nomenclature, although formally they differ. In particular, the $\wedge$-sign of $\hat{\u}$, $\hat{\v}$, $\hat{\w}$, $\hatgradc$, $\hatdiv$ and $\hatlaplace$ will be omitted in the following.

\subsection{Time discretisation}

The procedure applied for the discretisation of time is based on \citet{chorin} and \citet{onchorin} and was successfully applied by \citet{Reuther_PF_2018} for the surface NS equation. For numerical reasons, we linearise the nonlinear covariant directional derivative and obtain $\u^*\cdot\gradc \u^* \approx \u^n\cdot\gradc \u^*$. This yields the following problem with a semi-implicit Euler time scheme:

Let $\u^0:=\u_0\in TM$ be the sufficiently smooth initial velocity field and $\tau_n\in\R$ be the time step in the $n$-th iteration. For $n\to n+1$ determine successively
\begin{enumerate}
    \item $\u^*,\v^*$ and $\w^*$ such that
    \begin{eqnarray}
       \hspace*{-0.4cm} \frac{1}{\tau_n}(\u^*\!-\!\u^n)+(\u^n\!\cdot\!\gradc)\u^* \!- \Y_0 \v^* + \Y_2 \w^*
        \!- \Y_4(\laplace \w^* \!+ \kappa \w^*) + \alpha(\nub\cdot \u^*)\nub &=&0, \label{eq:time algorithm active ns}\\[-0.5em]
        \v^* \! -\laplace \u^* \!- \kappa \u^* + \alpha(\nub\cdot \v^*)\nub&=&0, \label{eq:time algorithm v}\\
        \w^* \! -\laplace \v^* \!- \kappa \v^* + \alpha(\nub\cdot \w^*)\nub&=&0. \label{eq:time algorithm w}
    \end{eqnarray}
    \item $p^{n+1}$ such that
    \begin{equation}
        \tau_n\Delta_M p^{n+1}=\div \u^*, \label{eq:time algorithm poisson}
    \end{equation}
    \hspace*{0.3cm} with $\Delta_M$ the Laplace-Beltrami operator.
    \item $\u^{n+1}$ such that
    \begin{equation}
        \u^{n+1} = \u^*-\tau_n\cdot\gradc p^{n+1}. \label{eq:time algorithm projection}
    \end{equation}
\end{enumerate}

\subsection{Space discretisation}

A regular triangulation $M_h= \bigcup_{T\in\cal T\it}T$ of the smooth surface $M$ is constructed by triangular elements $T\in\cal T$ determined by fixed points that are distributed equally over the surface. Note that the particular choice of $M_h$ in general also affects the normal vector field. If analytically known, we define the unit normal vector field $\nub_h$ of $M_h$ as the analytic normal of $M$ in each degree of freedom (DOF). Otherwise, in order to achieve convergence, an approximation is needed which is at least one order better than the approximation of the surface \citep{Hansbo_IMAJNA_2020}. This can, e.g., be obtained by locally reconstructing higher order approximations of $M_h$ and computing the normals from them, as, e.g., considered in \citet{Reuther_JFM_2020,Nitschke_PRSA_2020}. 

The surface differential operators are manipulated to operate on $M_h$ instead of $M$ essentially by using the pointwise normal projection onto $M_h$, which is given by $\pimh=I-\nub_h\nub_h^T$. We apply a scalar-valued surface finite element method for each component of the partial differential equations \citep{Dziuk_AN_2013}. The procedure is analogous to the regular well-studied finite element method in flat space with the only difference of a surface discretisation \citep{Veyetal_CVS_2007}. The weak derivative, Sobolev spaces etc. can be defined in the same manner. We consider the piecewise linear finite element space $V_h=\left\{ \varphi\in L^2(M_h):\varphi|_T\in\mathbb{P}^1 \textrm{ for all } T\in\cal T\it \right\}$ for trial and test functions. For $k=1,2,3$ let $\uk$, $\vk$, $\wk$, $\unk$, $\vnk$ and $\wnk$ be the sufficiently smooth $k$-th component of $\u^*$, $\v^*$, $\w^*$, $\u^n,\v^n$ and $\w^n$, respectively. We multiply each component of equations (\ref{eq:time algorithm active ns})-(\ref{eq:time algorithm projection}) with a smooth test function $\varphi\in V_h$, integrate over the domain $M_h$ and apply the divergence theorem to achieve the weak formulation. This yields the following time and space discrete problem:

Let $\u^0=\bracket{u^0_k}_{k}$ with $u^0_k\in V_h$ for all $k=1,2,3$ be the initial velocity field. For $n\to n+1$ determine successively
\begin{enumerate}
    \item $\uk,\vk,\wk\in V_h$ such that
    \begin{eqnarray}
        \frac{1}{\tau_n}\intmh{\uk\cdot\varphi} + \intmh{\u^n\gradcc\uk\cdot\varphi} - \Y_0\intmh{\vk\cdot\varphi}  \nonumber\\
        + \Y_2\intmh{\wk\cdot\varphi} + \Y_4\intmh{\gradcc\wk\cdot\gradcc\varphi} -\Y_4\intmh{\kappa\wk\cdot\varphi} \nonumber\\
        + \alpha\sum_{i=1}^3\intmh{\nu_i u^*_i\nu_k\cdot\varphi} = \frac{1}{\tau_n}\intmh{\unk\cdot\varphi},
    \end{eqnarray}
    \begin{eqnarray} \label{eq:final weak formulation v}
        \intmh{\vk\cdot\varphi} &+& \intmh{\gradcc\uk\cdot\gradcc\varphi} - \intmh{\kappa\uk\cdot\varphi} \nonumber\\
        &+& \alpha\sum_{i=1}^3\intmh{\nu_i v^*_i\nu_k\cdot\varphi} =0,
    \end{eqnarray}
    \begin{eqnarray} \label{eq:final weak formulation w}
        \intmh{\wk\cdot\varphi} &+& \intmh{\gradcc\vk\cdot\gradcc\varphi} - \intmh{\kappa\vk\cdot\varphi} \nonumber\\
        &+& \alpha\sum_{i=1}^3\intmh{\nu_i w^*_i\nu_k\cdot\varphi} = 0
    \end{eqnarray}

    for every test function $\varphi\in V_h$ and for all $k=1,2,3$. Thereby, $\u^{n}=\bracket{u^{n}_k}_{k}$ with $u^{n}_k\in V_h$ for all $k=1,2,3$ denotes the solution from the previous time step $n$.
    
    \item $p^{n+1}\in V_h$ such that
    \begin{equation} \label{eq:weak relaxation scheme}
        \tau_n\intmh{\gradcc p^{n+1}\cdot\gradcc\varphi} + \intmh{\div \u^*\cdot\varphi} = 0
    \end{equation}
    for every test function $\varphi\in V_h$.
    
    \item $u^{n+1}_k\in V_h$ such that
    \begin{equation}
        \intmh{u^{n+1}_k\cdot\varphi} = \intmh{u^*_k\cdot\varphi}-\tau_n\intmh{\left[\gradcc p^{n+1}\right]_k\cdot\varphi}. \label{eq:final time algorithm projection}
    \end{equation}
    for every test function $\varphi\in V_h$ and for all $k=1,2,3$.
\end{enumerate}
The implementation of this algorithm is done with the help of the C++ based finite element library AMDiS \citep[see][]{Veyetal_CVS_2007, Witkowskietal_ACM_2015} and the resulting assembled linear systems are solved by BiCGSTAB($\ell$) \citep{bicstab3}.

\subsection{Validation}\label{sec:validation}

We validate the numerical approach against known results for passive flows on a torus and active flows on a sphere. In the following, all presented numbers and values are dimensionless quantities.

\subsubsection{Passive flows on torus}

\begin{figure}
    \centering
    \includegraphics[width=.5\linewidth]{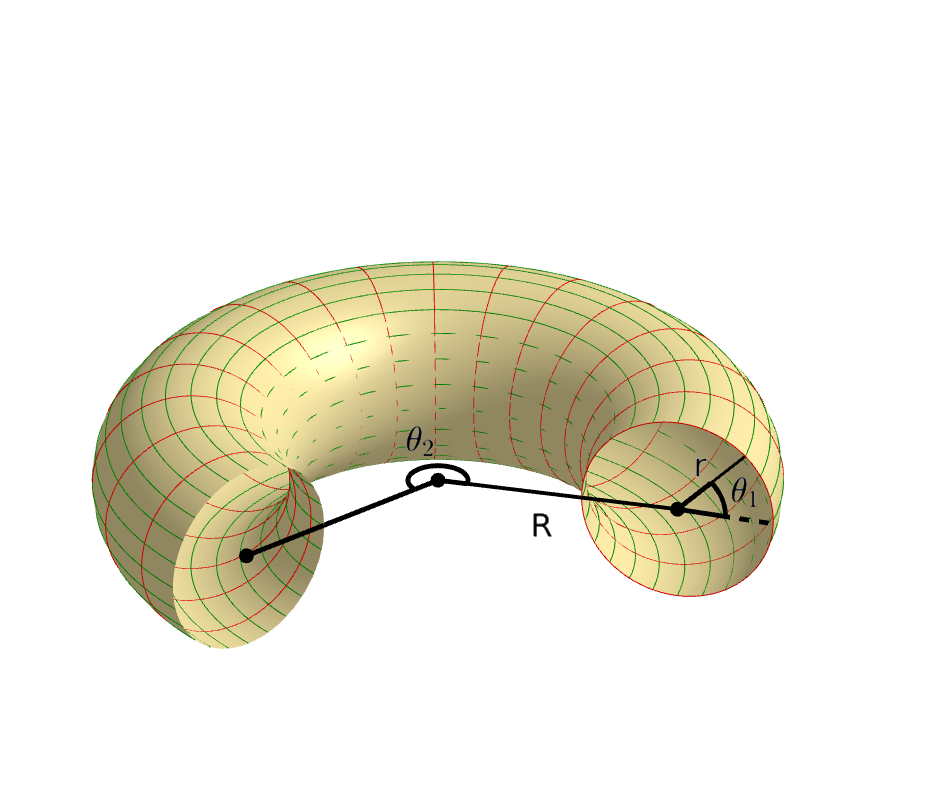}
    \caption{Parameterisation of torus with inner and outer radius $r$ and $R=1/r$, and corresponding angles $\theta_1$ and $\theta_2$.}
    \label{fig:torus parametrization}
\end{figure}

\begin{figure}
    \begin{minipage}[t]{.033\textwidth}
        (a)
    \end{minipage}%
    \begin{minipage}[t]{.45\textwidth}
        \textrm{}
        \centering
        \vspace{0pt}
        \adjincludegraphics[valign=t,width=0.9\textwidth]{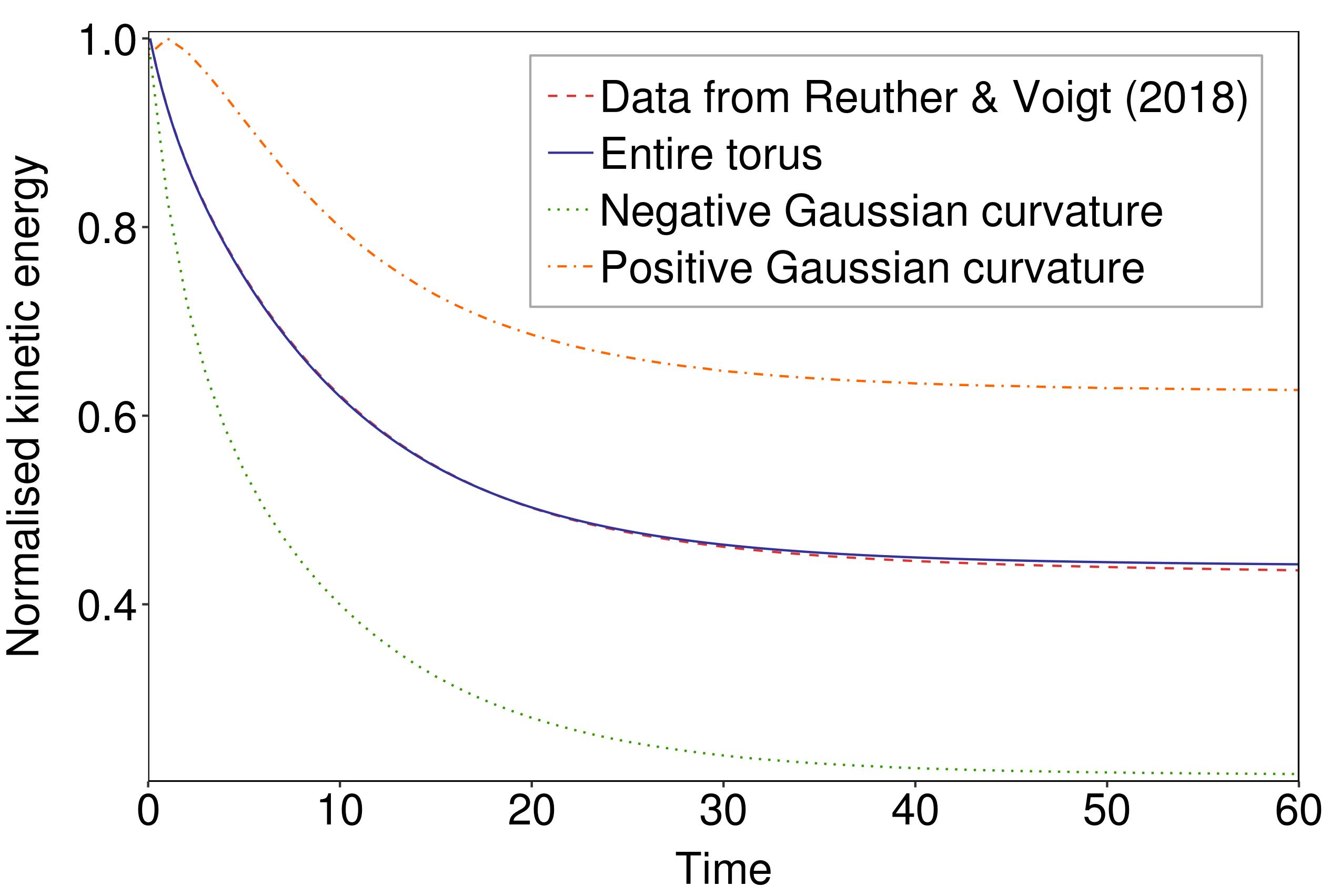}
    \end{minipage}%
    \hfill
    \begin{minipage}[t]{.033\textwidth}
        (b)
    \end{minipage}%
    \begin{minipage}[t]{.45\textwidth}
        \textrm{}
        \centering
        \vspace{0pt}
        \adjincludegraphics[valign=t,width=0.9\textwidth]{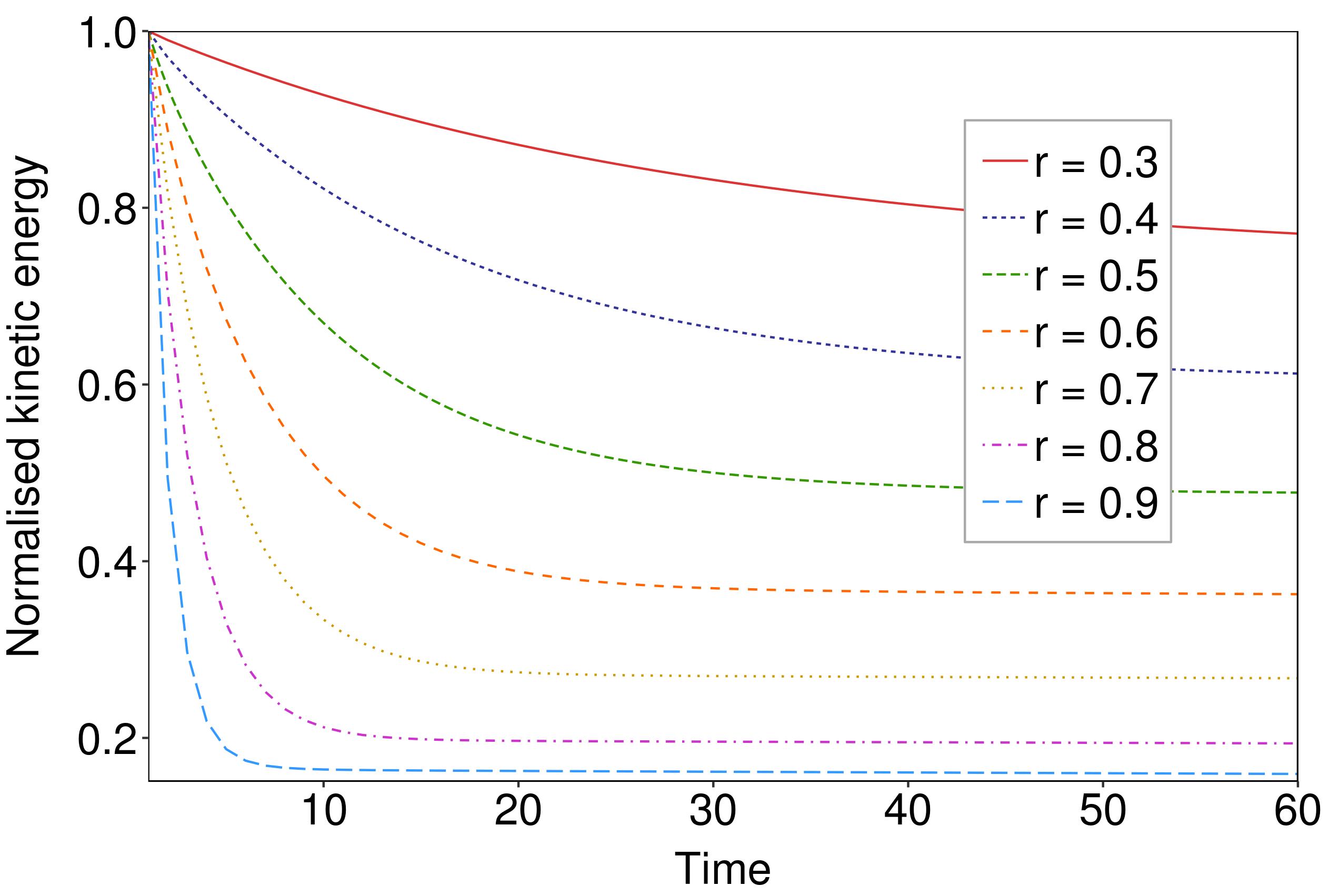}
    \end{minipage}
    \caption{Normalised total kinetic energy of the implemented model with parameters $\Y_0=0.1$ and $\Y_2,\Y_4=0$, penalty parameter $\alpha=3000$, time step $\tau_n=0.1$ and
    (a) torus radii $r=0.5$ and $R=2.0$ and the results from \citet{Reuther_PF_2018} (dashed line), on the entire surface and on regions of positive and negative Gaussian curvature, and
    (b) different choices of radii $r=0.3,0.4,...,0.9$, with $R=1/r$ accordingly, on the entire domain, against time. The distinction between regions of positive and negative Gaussian curvature for different $r$ leads to the same behaviour as shown in (a).}
    \label{fig:energy diagram torus non active}
\end{figure}

We first consider passive flows on a torus. This has been done in detail by \citet{Reuther_PF_2018}, utilising a surface NS equation which is achieved by $\Y_0 = 1/\textrm{Re}$ and $\Y_2=\Y_4=0$. We consider the time step $\tau_n=0.1$ and the penalty parameter $\alpha=3000$. The used mesh of the torus surface with outer radius $R=2.0$ and inner radius $r=0.5$ consists of 49.280 triangular elements and 24.640 vertices. To set the initial velocity field $\u_0$, we consider the arithmetic mean of the two linear independent harmonic vector fields, $$u_{\theta_1}^{harm} = \frac{1}{2 \|\mathbf{x}\|_2} \partial_{\theta_1} \mathbf{x}, \qquad u_{\theta_2}^{harm} = \frac{1}{4 \|\mathbf{x}\|_2^2} \partial_{\theta_2} \mathbf{x},$$ 
with $\theta_1, \theta_2 \in [0, 2 \pi]$, see Figure \ref{fig:torus parametrization} for a definition of the parameters. We normalise the total kinetic energy of the system $E(t)=\frac{1}{2}\integrate{M_h}{\|\u(t)\|^2}{A}$ by dividing by the maximum value $E_{max}=E(0)$ and compare it with the benchmark problem in \citet{Reuther_PF_2018}, see Figure \ref{fig:energy diagram torus non active}(a). The agreement gives a first validation of the numerical approach. Figure \ref{fig:energy diagram torus non active}(b) shows the normalised kinetic energy over time for different choices of the inner radius $r$ and according outer radius $R=1/r$ to preserve the surface area of the torus. Already for this case a dependency of the dynamics on geometric properties is realised. This includes not only a faster decay for increasing inner radius $r$, but also a faster decay in regions of negative Gaussian curvature, see Figure \ref{fig:energy diagram torus non active}(a).

\subsubsection{Active flows on sphere} \label{section: Active flows on sphere}

\begin{figure}
    \begin{minipage}[t]{.033\textwidth}
        (a)
    \end{minipage}%
    \begin{minipage}[t]{.248\textwidth}
        \textrm{}
        \centering
        \vspace{0pt}
        \adjincludegraphics[valign=t,width=0.9\textwidth]{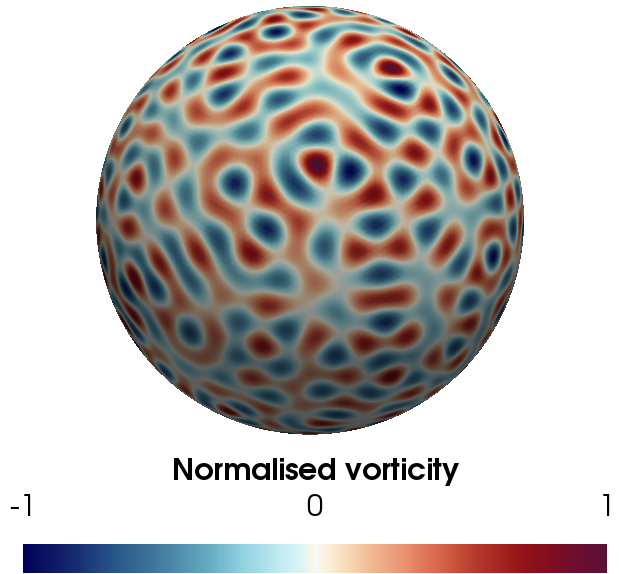}
    \end{minipage}%
    \begin{minipage}[t]{.033\textwidth}
        (b)
    \end{minipage}%
    \begin{minipage}[t]{.24\textwidth}
        \textrm{}
        \centering
        \vspace{0pt}
        \adjincludegraphics[valign=t,width=0.9\textwidth]{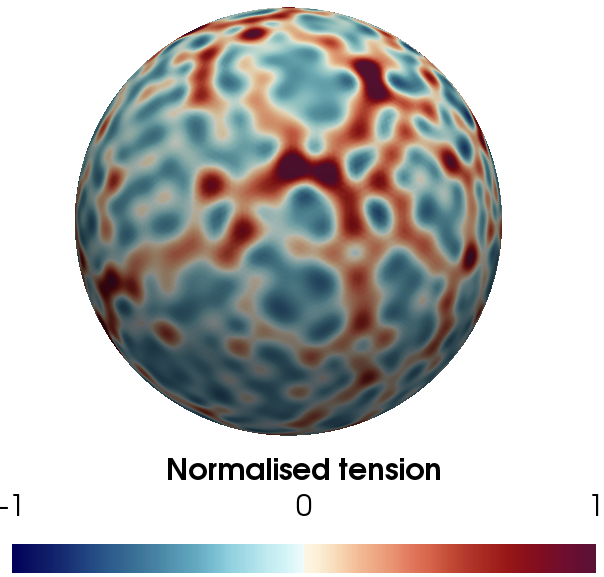}
    \end{minipage}%
    \begin{minipage}[t]{.033\textwidth}
        (c)
    \end{minipage}%
    \begin{minipage}[t]{.42\textwidth}
        \textrm{}
        \centering
        \vspace{0pt}
        \adjincludegraphics[valign=t,width=0.9\textwidth]{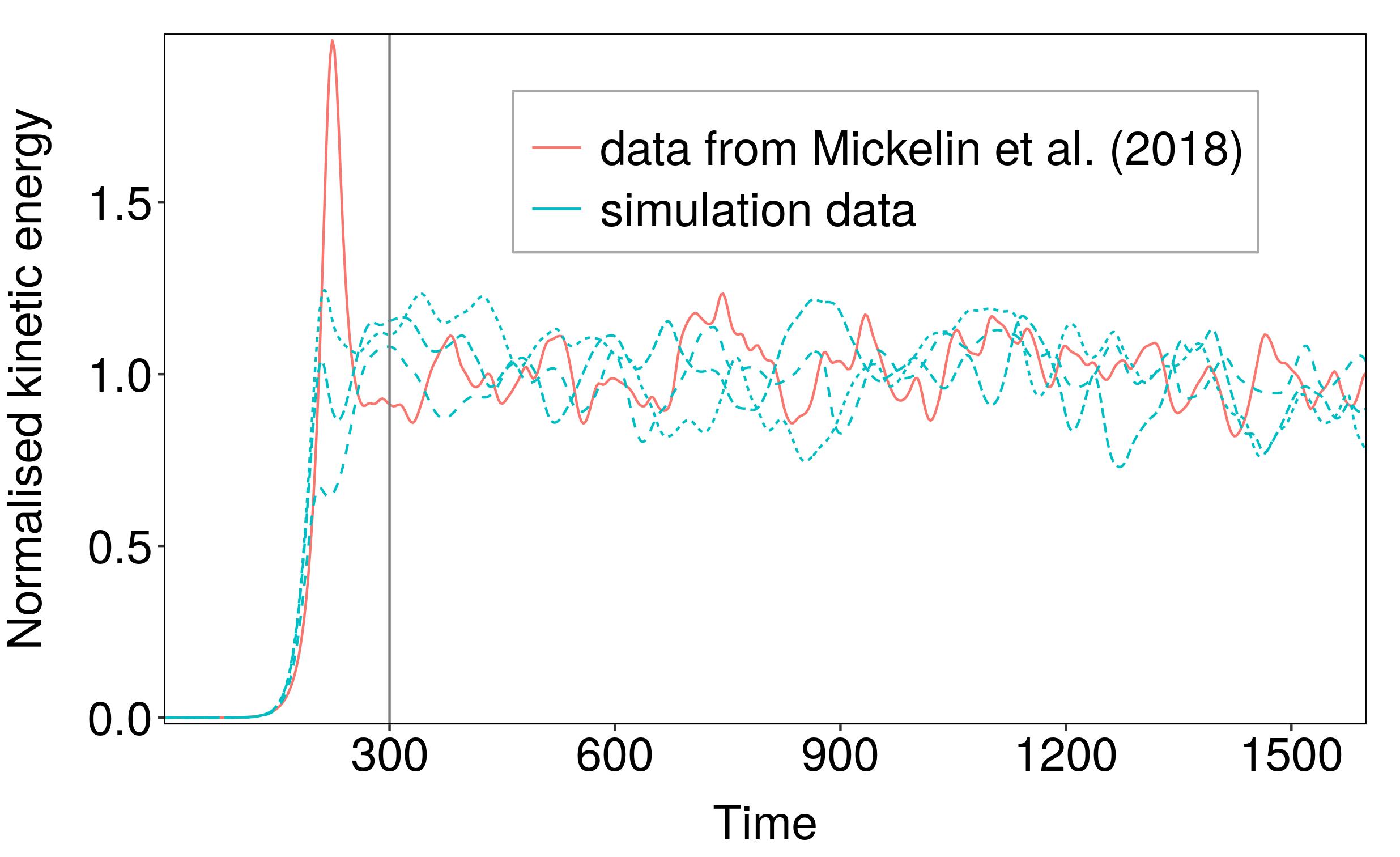}
    \end{minipage} \\
    \begin{minipage}[t]{.033\textwidth}
        (d)
    \end{minipage}%
    \begin{minipage}[t]{.95\textwidth}
        \textrm{}
        \centering
        \vspace{0pt}
        \adjincludegraphics[valign=t,width=0.9\textwidth]{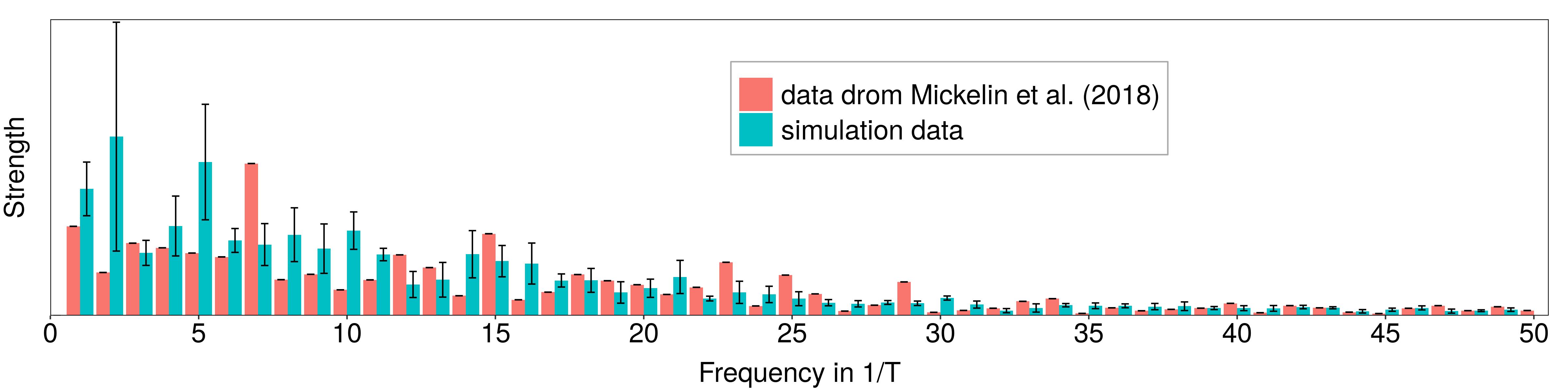}
    \end{minipage}
    \caption{Snapshots of (a) vorticity and (b) surface tension field at time $t=860$. (c) Normalised kinetic energy - total kinetic energy divided by mean kinetic energy after relaxation, indicated by vertical grey line at $t = 300$ - against time in comparison with data from \citet{Mickelin_PRL_2018}. Considered parameters are $\Y_0=3.901565\cdot10^{-2}, \Y_2=-7.886719\cdot10^{-5}$ and $\Y_4=3.887\cdot10^{-8}$. Different line types emerge from different random initial conditions. (d) Distribution of frequencies found by FFT averaged over three simulation runs, the strength is given by the absolute value of the according coefficient of the FFT. Black lines indicate the according sample standard deviation.}
    \label{fig:energy diagram sphere}
\end{figure}

We next compare active flows on a sphere with results in \citet{Mickelin_PRL_2018}. We consider parameters leading to anomalous vortex-network turbulence, compare case (d) in Figure 2 of \citet{Mickelin_PRL_2018}.
The penalty parameter $\alpha=3,000$ is as in the passive case, but the time step is reduced to $\tau_n=0.005$. As initial velocity field we consider a random field. The sphere is discretised by 50,000 triangular elements and 25,002 vertices. The radius of the sphere is $R = 1$. Figure \ref{fig:energy diagram sphere}(a) shows a snapshot of the normalised vorticity field. The vorticity is computed by $\phi = \curl \u$, with $\curl$ the surface rotation, defined by $\curl \u = \div (\u \times \nub)$ and $\u$ to be interpreted as the tangential velocity field supplemented with a third component in normal direction set to be zero. $\phi$ is a scalar field, the magnitude of the vector pointing in normal direction. Figure \ref{fig:energy diagram sphere}(b) shows the corresponding normalised surface pressure/surface tension. To be comparable the considered normalisation follows \citet{Mickelin_PRL_2018}. Anomalous vortex-network turbulence is characterised in \citet{Mickelin_PRL_2018} by topological measures of the vorticity and geometric measures of the high-tension domains. The last shows highly branched chain-like structures which are clearly visible in Figure \ref{fig:energy diagram sphere}(b). This provides a further qualitative validation of the numerical approach. Finite-size vortices self-organise into chain complexes of antiferromagnetic order that percolate trough the entire surface forming an active dynamic network, which provides an efficient mechanism for upward energy transport from smaller to larger scales. The dynamics of this process is validated against \citet{Mickelin_PRL_2018} in Figure \ref{fig:energy diagram sphere}(c) by comparing the normalised kinetic energy, the total kinetic energy divided by the mean kinetic energy after relaxation. As the initial data in \citet{Mickelin_PRL_2018} is not known, we consider different simulations with different initial data. The relaxation time, as well as the amplitude and frequency of the observed oscillations are independent on the initial data and in reasonable agreement with \citet{Mickelin_PRL_2018}. To compare the amplitudes of the energy curves, we have computed the root mean square $\textrm{RMS} = (\frac{1}{T_2-T_1}\int_{T_1}^{T_2} \left(E(t)-\bar{E}\right)^2 \, \mathrm{d}t)^\frac{1}{2}$ with $T_1=300$ and $T_2=1600$, and mean kinetic energy $\bar{E}$. The computed values are $\textrm{RMS}=0.129, 0.109$ and $0.070$ for our simulations and $\textrm{RMS}=0.087$ for the data of \citet{Mickelin_PRL_2018} with peak-to-peak amplitude values of $0.487, 0.486, 0.333$ and $0.415$, respectively. A Fast Fourier Transform (FFT) gives the dominant frequencies, see Figure \ref{fig:energy diagram sphere}(d). Quantitative differences result from different initial conditions but probably also from different approximations of geometric quantities. Other measures, such as size of vortices and high-tension domains are in agreement and qualitatively, our simulations are within the anomalous vortex-network turbulence regime identified in \citet{Mickelin_PRL_2018}. 

\begin{figure}
    \begin{minipage}[t]{.33\textwidth}
        \textrm{}
        \centering
        \vspace{0pt}
        \adjincludegraphics[valign=t,width=0.9\textwidth]{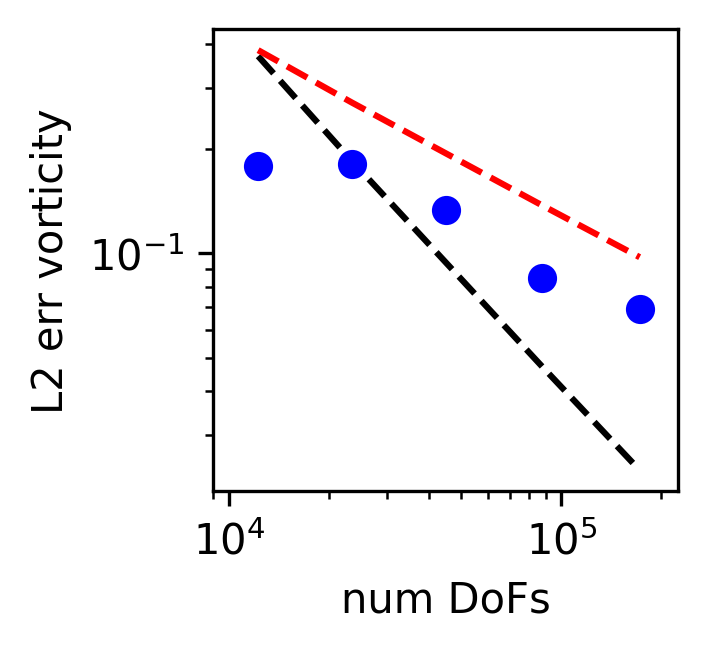}
    \end{minipage}%
    \begin{minipage}[t]{.33\textwidth}
        \textrm{}
        \centering
        \vspace{0pt}
        \adjincludegraphics[valign=t,width=0.9\textwidth]{results/sphere/L2Pressure_log.png}
    \end{minipage}%
    \begin{minipage}[t]{.33\textwidth}
        \textrm{}
        \centering
        \vspace{0pt}
        \adjincludegraphics[valign=t,width=0.9\textwidth]{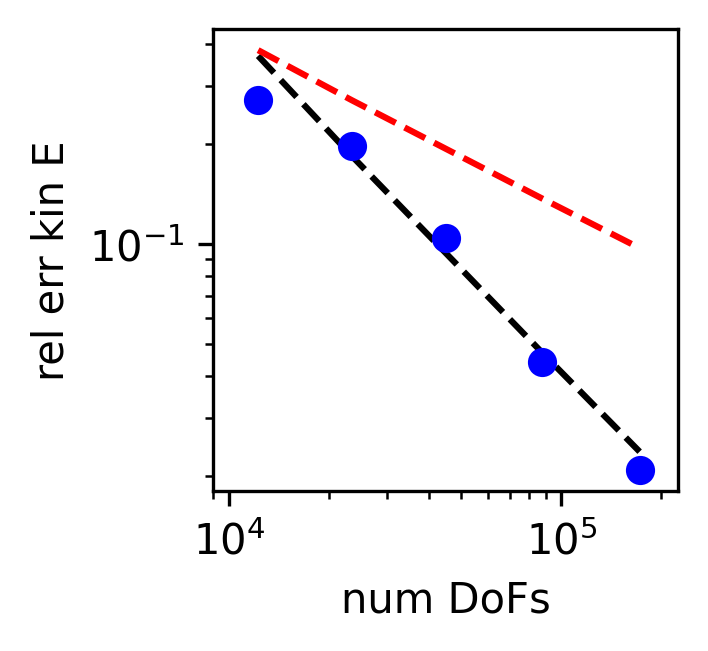}
    \end{minipage} 
    \caption{Convergence properties for vorticity, pressure/tension and kinetic energy. The $L^2$ error is computed with respect to the solution on the finest mesh and shown with respect to the number of degrees of freedom (DoFs). The black and the red lines indicate order $1$ and $1/2$, respectively.}
    \label{fig:convergence}
\end{figure}

To demonstrate the appropriate choice of the numerical parameters we consider different mesh resolutions. Mesh refinement is done by bisection with new vertices projected to the spherical surface. To circumvent relaxation effects in the initialisation phase we start with initial conditions, obtained from the previous simulations after $t = 300$, which are interpolated to the new meshes. The solutions are almost indistinguishable by eye from the one plotted in Figure \ref{fig:energy diagram sphere}(a) and (b). Figure \ref{fig:convergence} shows the $L^2$-errors in the vorticity $\phi$ and surface pressure / surface tension $p$, as well es differences in the kinetic energy $E$ after 400 time steps with constant time step $\tau_n$. The mesh used for the simulations shown in Figure \ref{fig:energy diagram sphere}(a) and (b) corresponds to the second data point in Figure \ref{fig:convergence}. The results indicate convergence of order $1/2$ with respect to the number of degrees of freedom (DoFs) for the vorticity. This relates to order $1$ with respect to mesh size. Surface pressure / surface tension and kinetic energy show convergence of order 1 with respect to number of DoFs and order 2 with respect to mesh size.    

These results for the full model on a sphere together with the results of the surface NS equation on a torus
make the proposed numerical scheme trustworthy to be applied for the full model on toroidal surfaces, in oder to explore the influence of local variations in Gaussian curvature on the anomalous turbulence regime.

\section{Results}\label{sec:results}

\subsection{Measures for topological and geometric quantities}

In analogy to \citet{Mickelin_PRL_2018} we determine the topology of the vorticity fields and the geometry of the high-tension
domains to identify anomalous turbulence. We therefore define the normalised Betti number as
$$
\mbox{Betti}_\phi(r) = \frac{\langle N_\phi(r,t) - N_\textrm{\tiny sphere}(t) \rangle}{\langle N_\textrm{\tiny sphere}(t) \rangle},
$$
with $N_\textrm{\tiny sphere}(t)$ denoting the zeroth Betti number of a sphere with radius $R=1$ at time $t$ as a reference value. The zeroth Betti number $N_\phi(r,t)$, with $r$ the inner radius of the considered torus, measures the number of connected domains with a high absolute vorticity $\{\x\in M:\phi(\x,t) > \alpha_\phi\cdot\max_{\x\in M}\phi(\x,t) \textrm{ or } \phi(\x,t) < \alpha_\phi\cdot\min_{\x\in M}\phi(\x,t)\}$ and 
a threshold $\alpha_\phi > 0$ at time $t$. The time average $\langle \cdot \rangle$ is taken after the initial relaxation period. Intuitively, large values of Betti$_\phi$ indicate many vortices of comparable circulation or many connected structures of similar size, whereas small values suggest the presence of a few dominant eddies or large connected structures. Accordingly, the normalised Branch number is defined by
$$
\mbox{Branch}_p(r) = \frac{\langle A_p(r,t) - A_\textrm{\tiny sphere}(t) \rangle}{\langle A_\textrm{\tiny sphere}(t) \rangle},
$$
with $A_\textrm{\tiny sphere}(t)$ serving as a reference value. The Branch number $A_p(r,t)$ denotes the mean of the ratios $\partial A/A$, where $A$ denotes the area and $\partial A$ denotes the boundary length of each connected component of the regions $\{\x\in M:p(\x,t) > \beta_p \Bar{p}(t)\}$ with 
$\Bar{p}$ its mean value and $\beta_p > 0$ a parameter. The ratio $\partial A/A$ is a measure of chainlike structures in the surface pressure/surface tension fields, 
a large value signaling a highly branched structure, whereas smaller values indicate less branching. $N_\textrm{\tiny sphere}(t)$ and $A_\textrm{\tiny sphere}(t)$ are considered to compare with anomalous turbulence in the spherical case in \citet{Mickelin_PRL_2018}.

\subsection{Active flows on toroidal surface}

We are using a mesh with 48,832 vertices and 97,664 resulting triangle elements, which we scale to approximate all tori with inner radius $r$ and according outer radius $R=1/r$. The total simulation time is $T=1600$, which turns out to be sufficient to reveal the characteristic active dynamics of the system.

\begin{figure}
    \begin{minipage}[t]{.033\textwidth}
        (a)
    \end{minipage}%
    \begin{minipage}[t]{.255\textwidth}
        \textrm{}
        \centering
        \vspace{0pt}
        \vfill
        \adjincludegraphics[valign=t,width=0.9\textwidth]{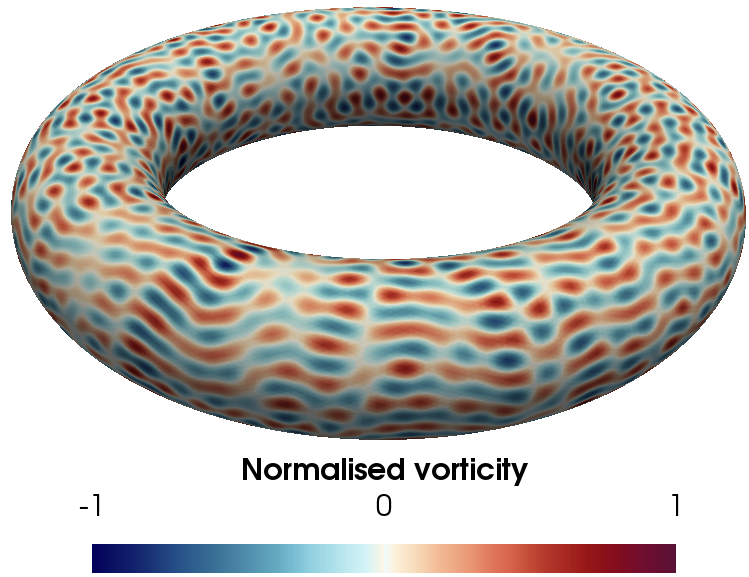}
    \end{minipage}%
    \begin{minipage}[t]{.033\textwidth}
        (b)
    \end{minipage}%
    \begin{minipage}[t]{.25\textwidth}
        \textrm{}
        \centering
        \vspace{0pt}
        \vfill
        \adjincludegraphics[valign=t,width=0.9\textwidth]{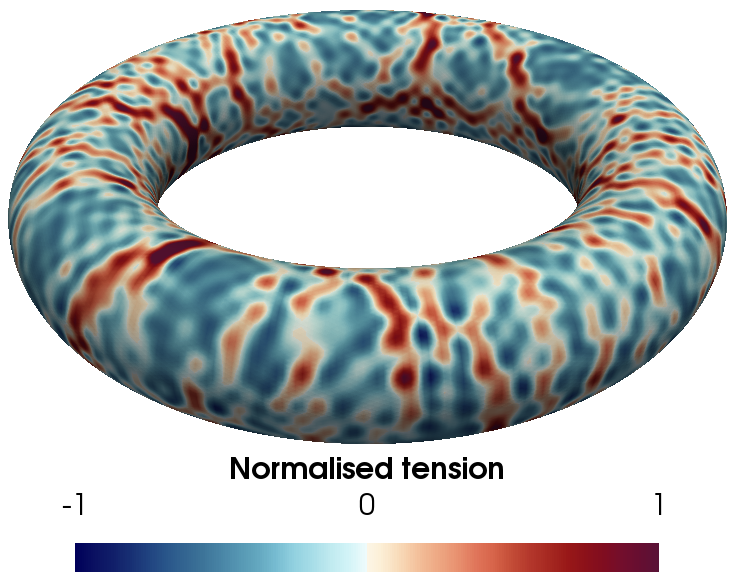}
    \end{minipage}%
    \begin{minipage}[t]{.033\textwidth}
        (c)
    \end{minipage}%
    \begin{minipage}[t]{.4\textwidth}
        \textrm{}
        \centering
        \vspace{0pt}
        \adjincludegraphics[valign=t,width=0.9\textwidth]{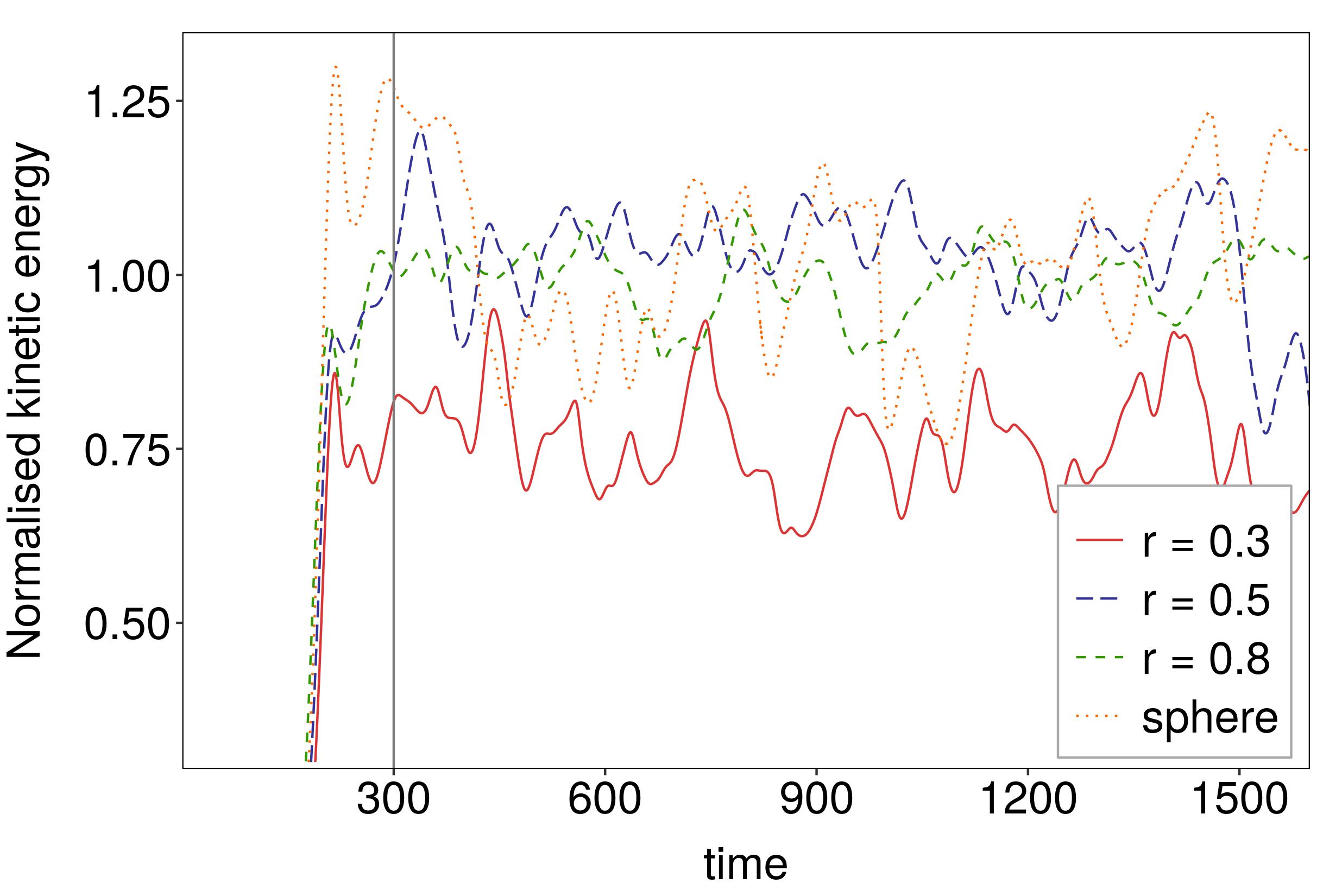}
    \end{minipage} \\
    \begin{minipage}[t]{.033\textwidth}
        (d)
    \end{minipage}%
    \begin{minipage}[t]{.95\textwidth}
        \textrm{}
        \centering
        \vspace{0pt}
        \adjincludegraphics[valign=t,width=0.9\textwidth]{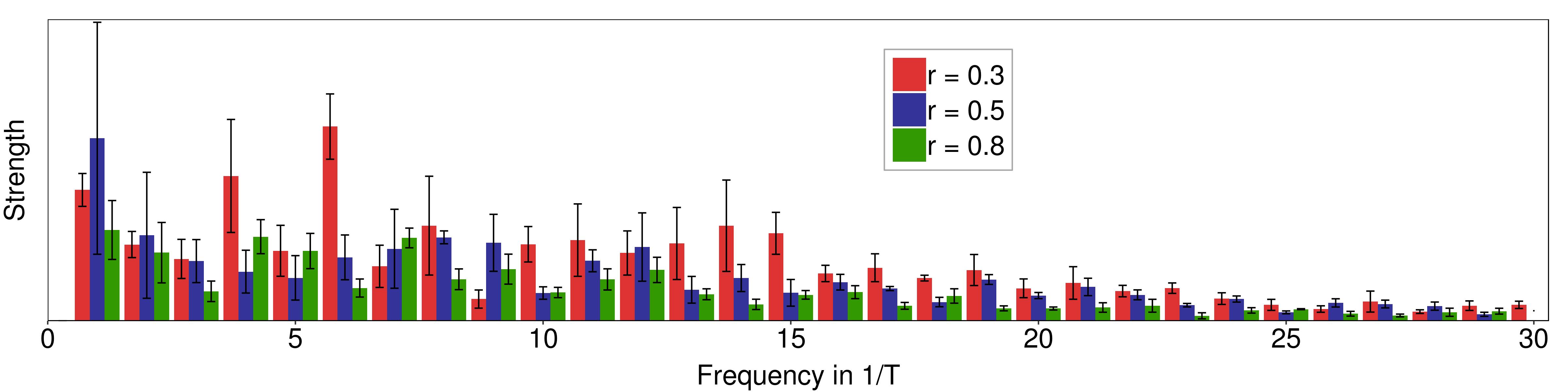}
    \end{minipage}
    \caption{Snapshots of (a) vorticity and (b) surface tension field on torus with radii $r=0.5$ and $R=2.0$. Other parameters are as in Figure \ref{fig:energy diagram sphere}. (c) Normalised kinetic energy - total kinetic energy for different choices of radii $r=0.3,0.5,0.8$ with $R=1/r$ divided by average total kinetic energy over all tori after relaxation - against time. The vertical grey line at $t = 300$ marks the beginning of the active turbulence regime. The results for a sphere are shown for comparison. (d) Distribution of frequencies found by FFT, averaged over three simulation runs, the strength is given by the mean absolute value of the according coefficient of the FFT. Black lines indicate the according sample standard deviation.}
    \label{fig:energy diagram torus active}
\end{figure}

The separation of the general model dynamics into a relaxation phase and an active chaotic phase can be done in the same way as already described for the sphere with generating and dissolving similar vortex patterns for all choices of inner radii $r$. Figure \ref{fig:energy diagram torus active}(a),(b) show snapshots of the normalised vorticity and the nomalised surface pressure/surface tension of a simulation on a torus. As for the sphere, the relaxation phase also runs until about $t=300$. Thereafter the active regime begins. Comparing the still images of the torus and the sphere, the vorticity seems to have the tendency to form slightly more complex, longer and maze-like structures on the torus. The tension still has the characteristic branched chain-like structures. However, the images give an impression on prefered orientations of the structure. Nevertheless, the patterns and dynamics appear to be qualitatively similar to the ones on a sphere. To investigate the differences between the simulation results on the sphere and the toroidal surfaces for different radii $r$ and $R=1/r$ we compute the normalised kinetic energy, see Figure \ref{fig:energy diagram torus active}(c), and the average RMS values over three simulation runs, see Table \ref{tab: rms}. The kinetic energy progressions of the different tori are similar to each other and to the sphere. Nevertheless, thin tori with inner radius $r=0.3,0.4$ appear to have larger energy fluctuations than thick tori with $r=0.8,0.9$. This observation is underlined by the peak-to-peak amplitude values. The distribution of frequencies in the FFT of the tori is shown in Figure \ref{fig:energy diagram torus active}(d). We can detect similar dominant frequencies for all tori with no significant differences between the various torus dimensions. Depending on the random initial conditions, distinct frequency distributions emerge, see Figure \ref{fig:energy diagram torus active}(d).

\begin{table}
    \centering
    \begin{tabular}{lcccccccc}
        Inner radius $r$ & 0.3 & 0.4 & 0.5 & 0.6 & 0.7 & 0.8 & 0.9 & sphere \\
        Average normalised kinetic energy & 0.761 & 0.855 & 1.032 & 1.080 & 1.059 & 0.992 & 0.931 & 1.017  \\
        Peak-to-peak amplitude & 0.494 & 0.316 & 0.347 & 0.220 & 0.227 & 0.222 & 0.260 & 0.504 \\
        RMS value & 0.093 & 0.061 & 0.067 & 0.044 & 0.041 & 0.052 & 0.051 & 0.128 \\
    \end{tabular}
    \caption{Average normalised kinetic energy, see Figure \ref{fig:energy diagram torus active}(c), peak-to-peak amplitude and root mean square (RMS) values of normalised kinetic energy curves, averaged over time and simulation runs for tori with inner radius $r$ and sphere with $R=1$.}
    \label{tab: rms}
\end{table}

To explain the observed differences in the energy fluctuations requires a deeper analysis of the data. The normalised Betti and Branch numbers for the whole surface have been computed for tori with inner radius $r=0.3,0.4,...,0.9$ and according outer radius $R=1/r$ with parameters $\alpha_\phi = 0.5$ and $\beta_p = 1$. They are divided by area and their averaged values, over different simulation runs, are given in Table \ref{tab: elongation}. We can observe lower normalised Betti numbers for very thin ($r=0.3$) and very thick ($r=0.9$) tori. The normalised Branch numbers show a different trend. They are largest for very thin ($r=0.3$) and smallest for thick ($r=0.8, 0.9$) tori.

\begin{table}
    \centering
    \begin{tabular}{lccccccc}
        Inner radius $r$ & 0.3 & 0.4 & 0.5 & 0.6 & 0.7 & 0.8 & 0.9 \\
        $\mbox{Betti}_\phi(r)$ per area& -0.436 & -0.161 & -0.166 & -0.067 & -0.055 & -0.182 & -0.568 \\
        $\mbox{Branch}_p(r)$ per area& 2.438 & 1.555 & 1.291 & 1.563 & 1.673 & 0.820 & 0.733 \\
    \end{tabular}
    \caption{Normalised Betti and Branch numbers per area for the entire surface averaged over simulation runs for tori with inner radius $r$.}
    \label{tab: elongation}
\end{table}

\begin{figure}
    \begin{minipage}[t]{.032\textwidth}
        (a)
    \end{minipage}%
    \begin{minipage}[t]{.19\textwidth}
        \textrm{}
        \centering
        \vspace{0pt}
        \adjincludegraphics[valign=t,width=\textwidth]{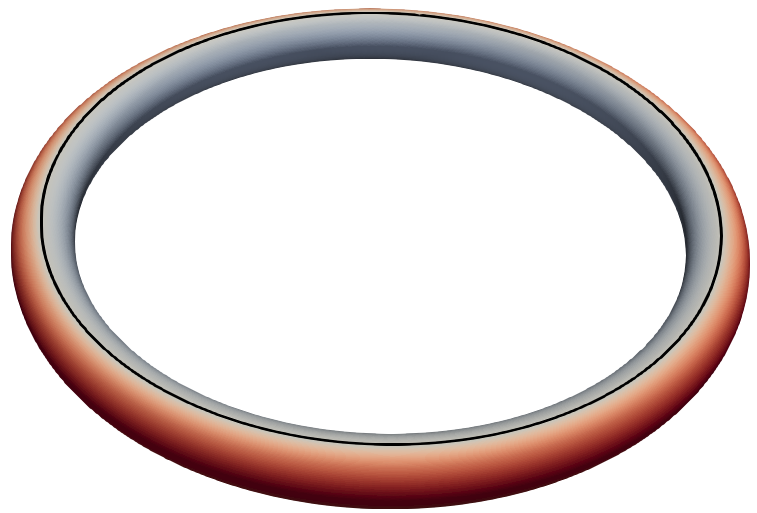}
    \end{minipage}%
    \hfill
    \begin{minipage}[t]{.032\textwidth}
        (b)
    \end{minipage}%
    \begin{minipage}[t]{.19\textwidth}
        \textrm{}
        \centering
        \vspace{0pt}
        \adjincludegraphics[valign=t,width=\textwidth]{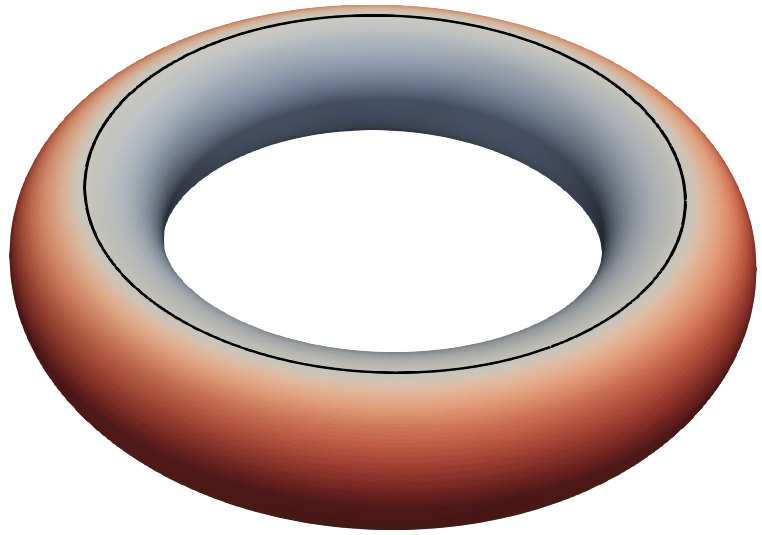}
    \end{minipage}
    \hfill
    \begin{minipage}[t]{.032\textwidth}
        (c)
    \end{minipage}%
    \begin{minipage}[t]{.19\textwidth}
        \textrm{}
        \centering
        \vspace{0pt}
        \adjincludegraphics[valign=t,width=\textwidth]{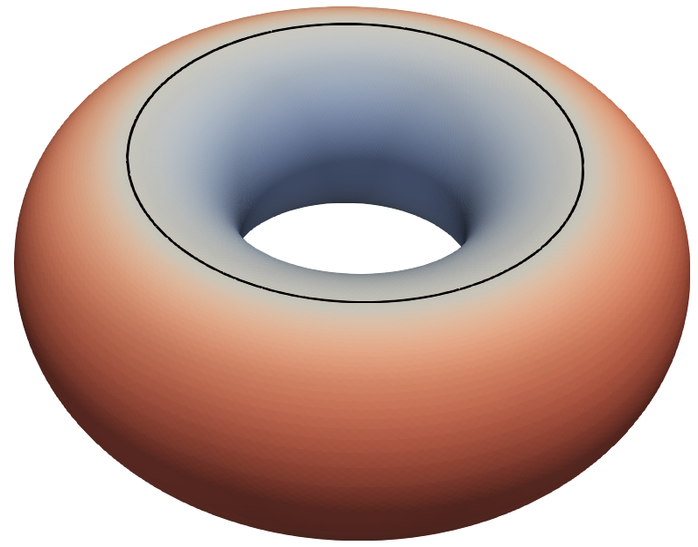}
    \end{minipage}%
    \hfill
    \begin{minipage}[t]{.032\textwidth}
        (d)
    \end{minipage}%
    \begin{minipage}[t]{.18\textwidth}
        \textrm{}
        \centering
        \vspace{0pt}
        \adjincludegraphics[valign=t,width=\textwidth]{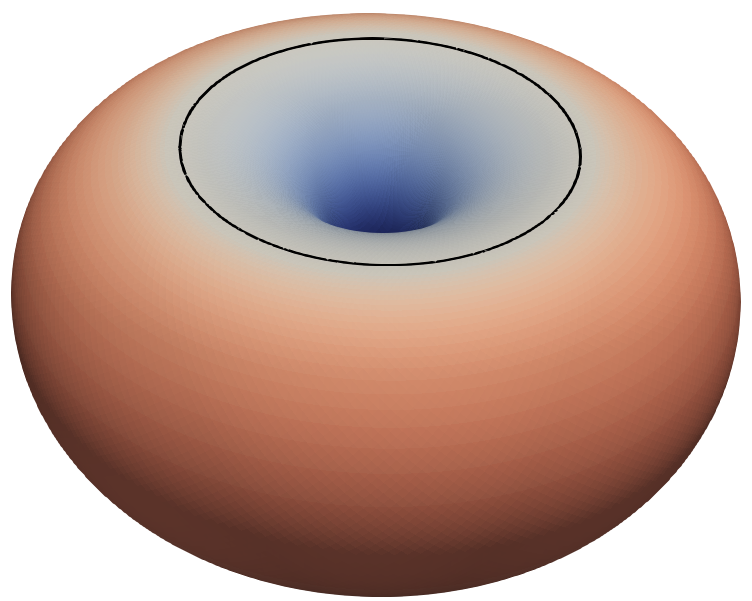}
    \end{minipage}
    \hfill
    \begin{minipage}[t]{.055\textwidth}
        \textrm{}
        \centering
        \vspace{0pt}
        \adjincludegraphics[valign=t,width=\textwidth]{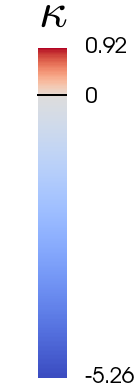}
    \end{minipage}
    \caption{Gaussian curvature $\kappa$ from blue to red on tori with inner radius a) $r=0.3$, b) $r=0.5$, c) $r=0.7$, d) $r=0.9$ and according outer radius $R=1/r$. $r$ does not scale between (a) - (d).}
    \label{fig:curvature tori}
\end{figure}

However, averaging over the entire surface does not account for the local differences in surface curvature. To extract a dependency of the normalised Betti and Branch numbers on local curvature we classify regions of equal Gaussian curvature on all considered tori. Figure \ref{fig:curvature tori} shows the different values of Gaussian curvature $\kappa$ on selected tori. While moderate values of positive Gaussian curvature are found on the outer part for all tori, strong negative values are only present in the inner part of thick tori ($r=0.8,0.9$). 

\begin{figure}
    \begin{minipage}[t]{.033\textwidth}
        (a)
    \end{minipage}%
    \begin{minipage}[t]{.42\textwidth}
        \textrm{}
        \centering
        \vspace{0pt}
        \adjincludegraphics[valign=t,width=\textwidth]{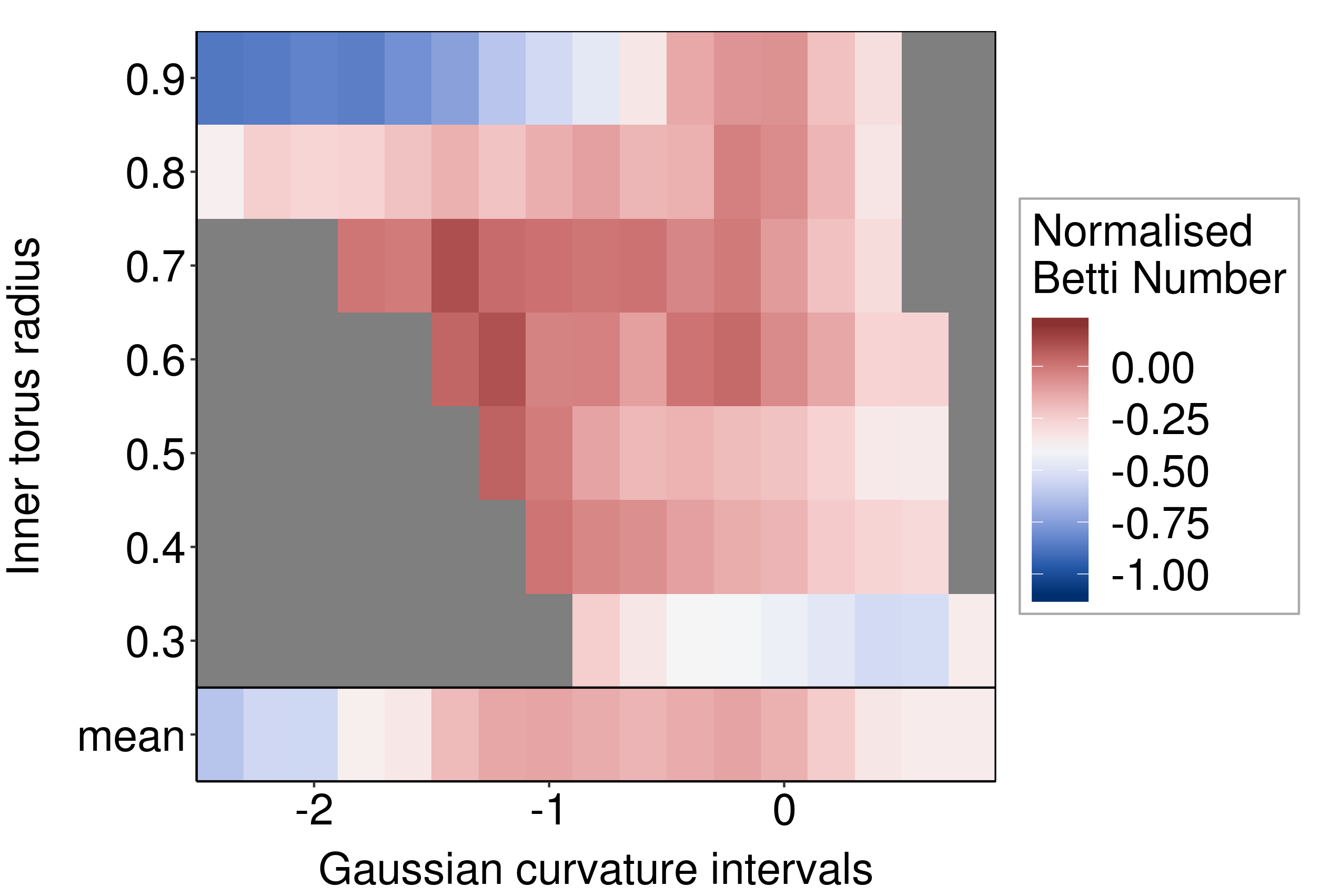}
    \end{minipage}%
    \hfill
    \begin{minipage}[t]{.033\textwidth}
        (b)
    \end{minipage}%
    \begin{minipage}[t]{.43\textwidth}
        \textrm{}
        \centering
        \vspace{0pt}
        \adjincludegraphics[valign=t,width=\textwidth]{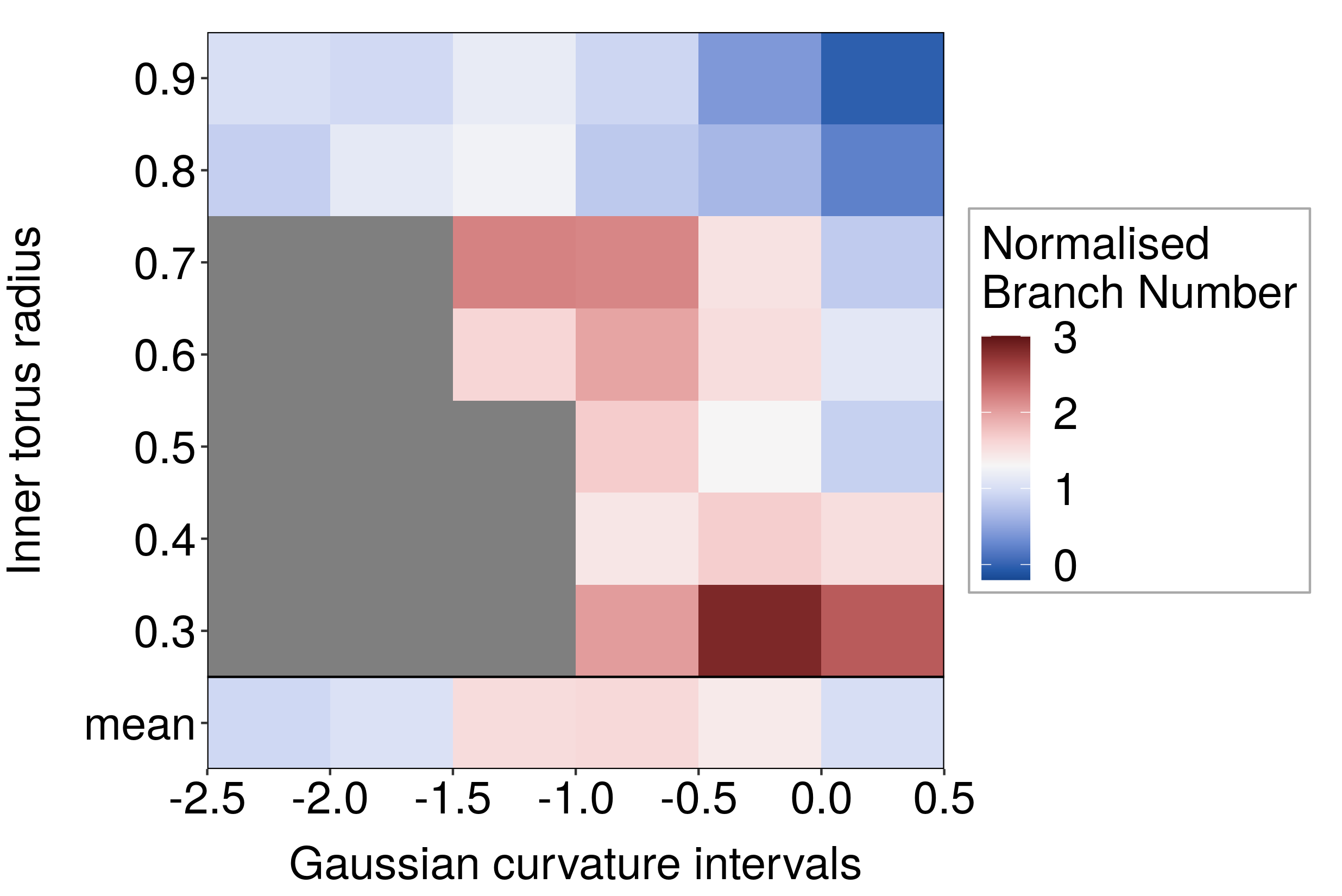}
    \end{minipage}
    \hspace*{\fill}\\
    \begin{minipage}[t]{.033\textwidth}
        (c)
    \end{minipage}%
    \begin{minipage}[t]{.46\textwidth}
        \textrm{}
        \centering
        \vspace{0pt}
        \adjincludegraphics[valign=t,width=.9\textwidth]{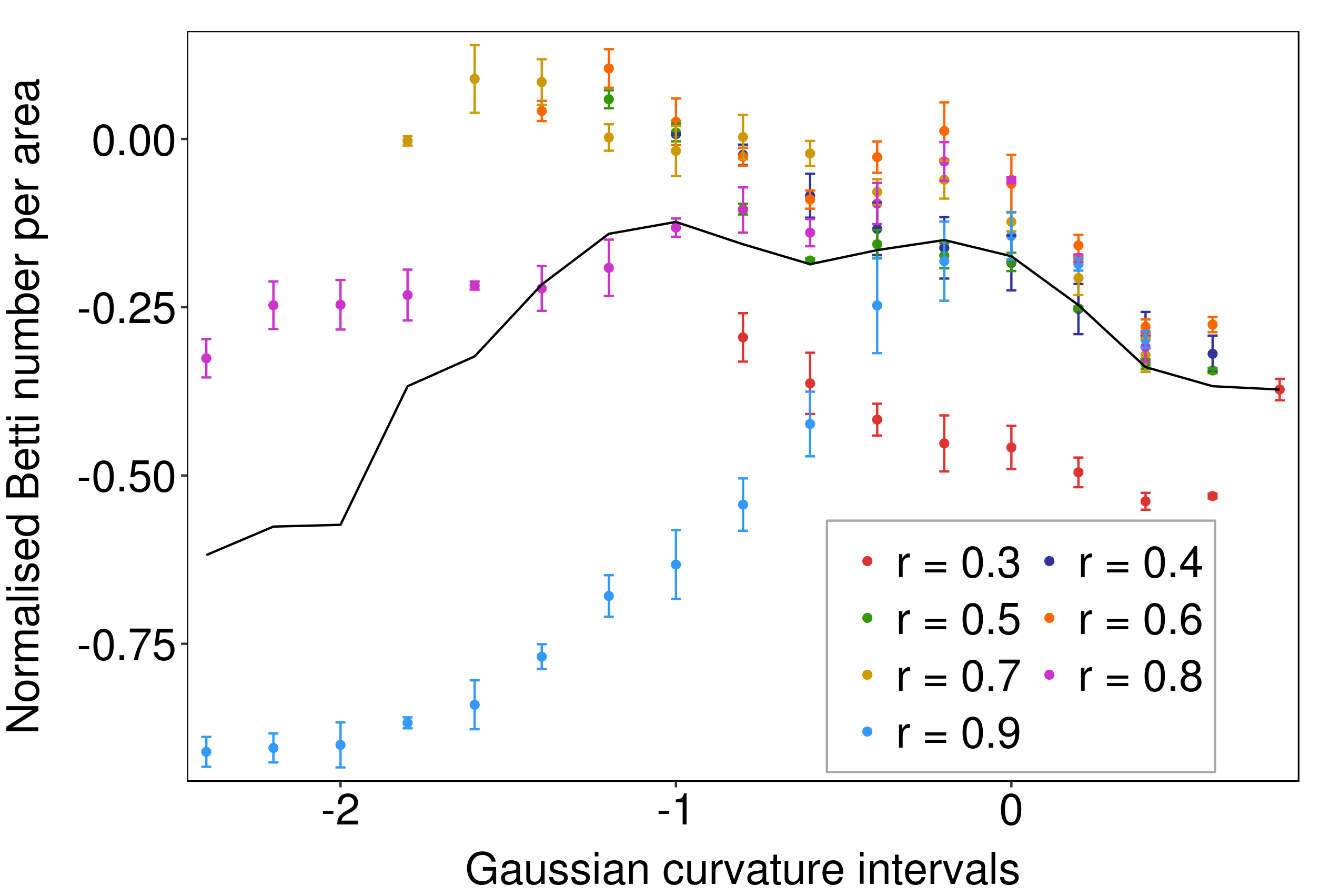}
    \end{minipage}%
    \hfill
    \begin{minipage}[t]{.033\textwidth}
        (d)
    \end{minipage}%
    \begin{minipage}[t]{.46\textwidth}
        \textrm{}
        \centering
        \vspace{0pt}
        \adjincludegraphics[valign=t,width=.9\textwidth]{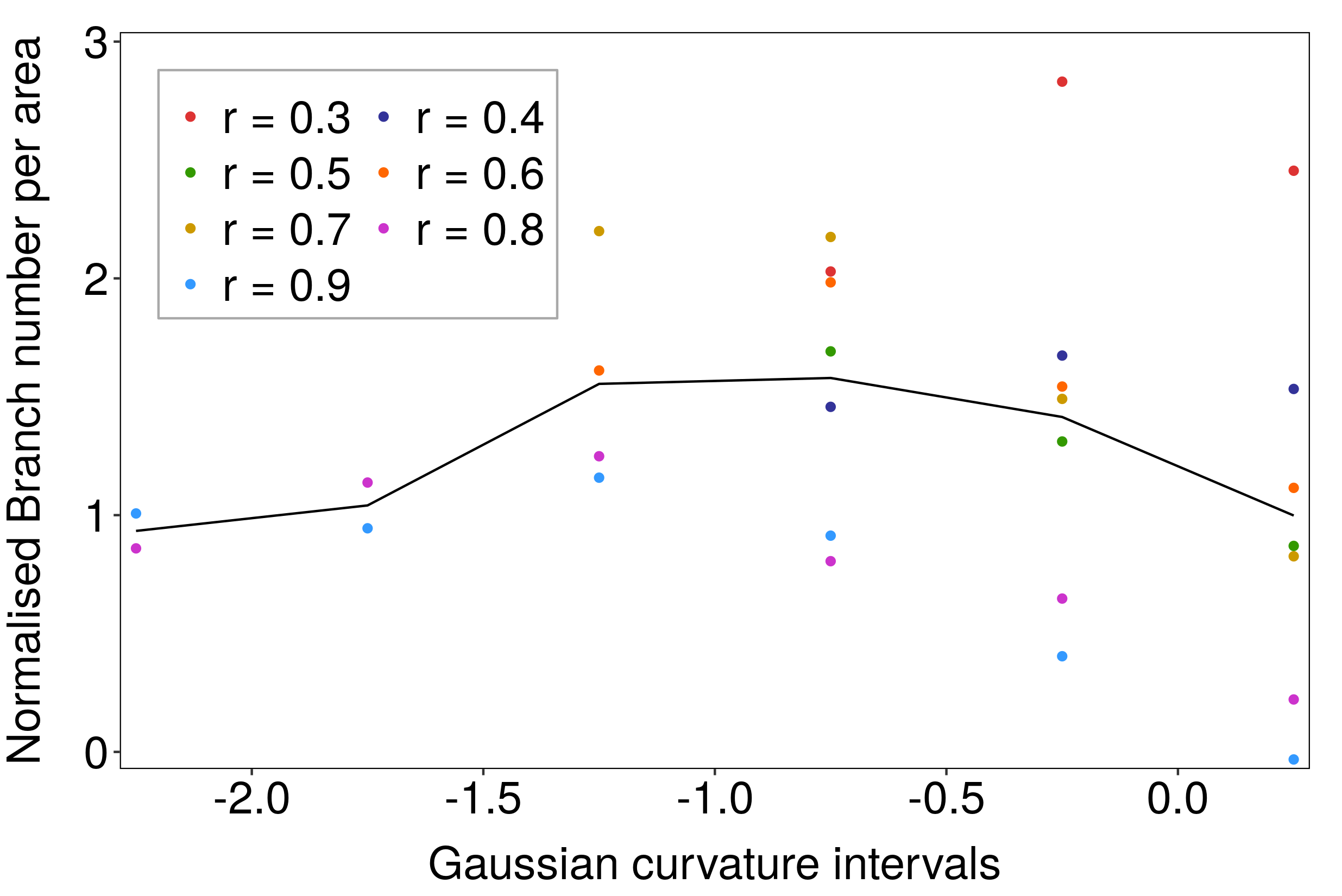}
    \end{minipage}
    \caption{a) Normalised Betti numbers per area for surface areas with Gaussian curvature $(\kappa-0.1,\kappa+0.1)$, $\kappa \in \{-2.4,-2.2,...,0.8\}$ and b) normalised Branch number for surface areas with Gaussian curvature $(\kappa-0.25,\kappa+0.25)$, $\kappa \in \{-2.25,-1.75,...,0.25\}$ vs. inner torus radius $r$, the row at the bottom showing the average value over all considered torus geometries. A few intervals of Gaussian curvature are not found on some tori, indicated by grey colour, see Figure \ref{fig:curvature tori}. The average values with according sample standard deviation of different simulation runs are shown in c) for the normalised Betti number and values for one simulation in d) for the normalised Branch number. The black lines show the according mean values over all runs and tori. The results are obtained with the parameters $\alpha_\phi = 0.5$ and $\beta_p = 1$. Results for different parameters are shown in the Appendix and demonstrate the robustness of the results.}
    \label{fig: Heatmaps}
\end{figure}

\begin{figure}
    \begin{minipage}[t]{.033\textwidth}
        (a)
    \end{minipage}%
    \begin{minipage}[t]{.42\textwidth}
        \textrm{}
        \centering
        \vspace{0pt}
        \adjincludegraphics[valign=t,width=\textwidth]{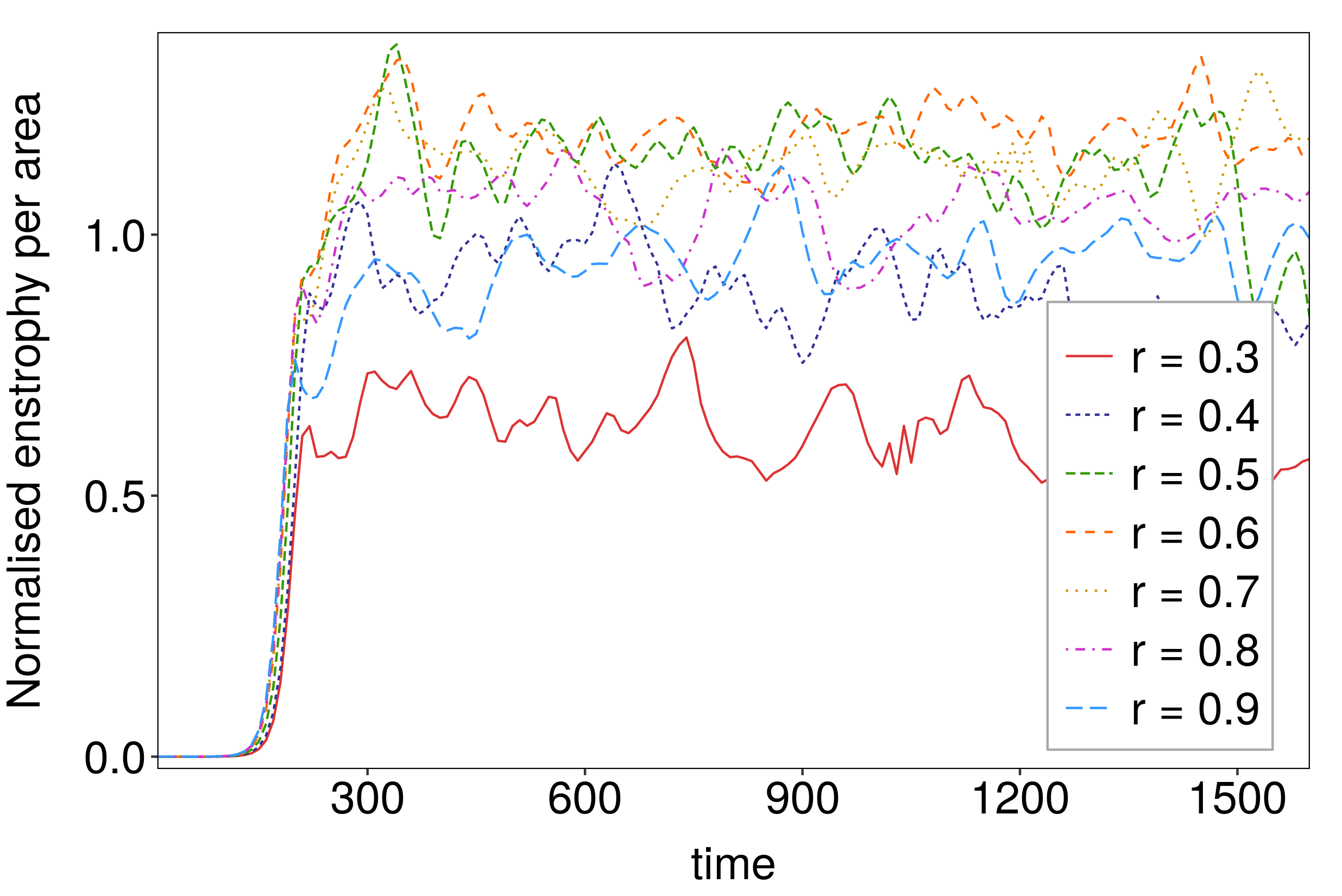}
    \end{minipage}%
    \hfill
    \begin{minipage}[t]{.033\textwidth}
        (b)
    \end{minipage}%
    \begin{minipage}[t]{.43\textwidth}
        \textrm{}
        \centering
        \vspace{0pt}
        \adjincludegraphics[valign=t,width=\textwidth]{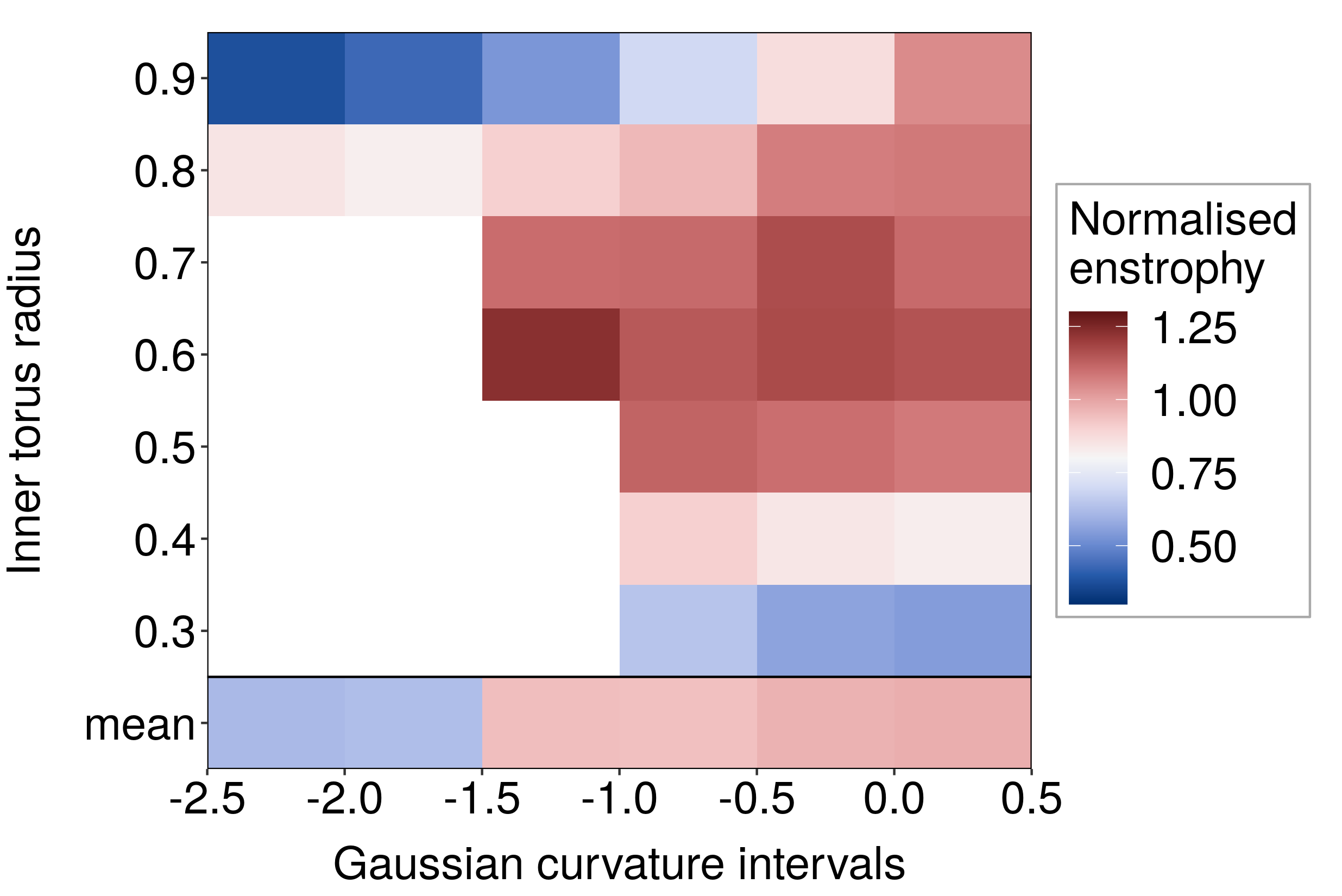}
    \end{minipage}
    \caption{a) Enstrophy per surface area nomalised by mean after relaxation over all tori, shown for inner torus radius $r = 0.3, \ldots, 0.9$. (b) Normalised enstrophy for surface areas with Gaussian curvature $(\kappa-0.25,\kappa+0.25)$, $\kappa \in \{-2.25,-1.75,...,0.25\}$ vs. inner torus radius $r$, the row at the bottom showing the average value over all considered torus geometries.}
    \label{fig:enstrophy}
\end{figure}

We visualise the normalised Betti and Branch numbers per area for the different tori according to intervals of Gaussian curvature, see Figure \ref{fig: Heatmaps}. This indicates a clear dependency of the normalised Betti number on the Gaussian curvature, with a maximal value for Gaussian curvature between $-1$ and $0$ and decreasing values for lower and higher Gaussian curvatures. Moderate values of Gaussian curvature show a similar behaviour as in the spherical case and thus indicate anomalous turbulence, whereas larger absolute values favour the presence of fewer more dominant vortices and thus deviation from the anomalous turbulence regime, see \citet{Mickelin_PRL_2018}. The behaviour for the normalised Branch number is similar, with maximal values for Gaussian curvature between $-1.5$ and $0$ and decreasing values for lower and higher Gaussian curvatures. While this is true for all tori, the values between the different tori strongly differ. They are largest for very thin tori ($r = 0.3$) and decrease towards very thick tori ($r = 0.9$). For all regions the normalised Branch number is larger than in the representative spherical case, indicating even enhance branching. The movies in the Electronic Supplement confirm this and show the characteristic percolation of the high tension structures through the entire domain. A characterisation of the flow regime as anomalous turbulence requires the topology of the vorticity fields and the geometry of the high-tension domains to be characteristic \citep{Mickelin_PRL_2018}. Both together thus indicate anomalous turbulence for moderate regions of Gaussian curvature, and possible deviations for lower and higher values. We also compute the enstrophy ${\cal{E}}(t) = \int_{M_h} \phi^2 d A$ for the whole surface and per region of constant Gaussian curvature, see Figure \ref{fig:enstrophy}(a) and (b), respectively. The values are largest for moderate tori ($r = 0.5, 0.6, 0.7$) and decay towards thinner ($r=0.3, 0.4$) and thicker ($r= 0.8, 0.9$) tori, see also Table \ref{tab:enstrophy}. 
In \citet{Mickelin_PRL_2018} it is argued that also the ratio between the mean kinetic energy and the mean enstrophy can be used to identify the anomalous turbulence regime. Even if this can only be a qualitative measure the ratio is shown in Table \ref{tab:enstrophy}. It shows a minimum for moderate tori ($r = 0.5, 0.6, 0.7$) and moderate Gaussian curvature values ($\kappa \in [-1.5, 0.0]$), for which anomalous turbulence is already identified and only slightly increases for thinner and thicker tori and lower and higher Gaussian curvature values, respectively. This weak indication on the influence of curvature effects on the active turbulence regime is consistent with the topological and geometric measures of the normalized Betti and Branch numbers.

\begin{table}
    \centering
    \begin{tabular}{lccccccc}
        Inner radius $r$ & 0.3 & 0.4 & 0.5 & 0.6 & 0.7 & 0.8 & 0.9 \\
        Normalised enstrophy per area & 0.632 & 0.893 & 1.138 & 1.195 & 1.142 & 1.049 & 0.954 \\
        $\langle E(t) \rangle / \langle {\cal{E}}(t) \rangle$ & 1.257 & 0.999 & 0.946 & 0.942 & 0.967 & 0.987 & 1.019 \\ \hline 
        Gaussian curvature $\kappa$ & -2.25 & -1.75 & -1.25 & -0.75 & -0.25 & 0.25 \\
        $\langle E(t) \rangle / \langle {\cal{E}}(t) \rangle$ & 1.089 & 1.078 & 0.881 & 0.967 & 0.974 & 1.034 \\
    \end{tabular}
    \caption{Mean normalised enstrophy per area over time and simulation runs for tori with inner radius $r$. Ratio of mean normalized kinetic energy and mean normalized enstrophy per area for tori with inner radius $r$ as well as intervals of Gaussian curvature $(\kappa-0.25,\kappa+0.25)$.}
    \label{tab:enstrophy}
\end{table}

For all considered measures we see a dependency on the global shape of the torus. However, the normalised Betti and Branch numbers, but also the nomalised enstrophy, indicate that the flow regime of anomalous turbulence depends also on the local geometry. Similar dependencies on local curvature have been found for topological defects in active nematic toroids \citep{Ellis_NP_2018}. These results are in contrast to equilibrium systems, which predict only a dependency on the global parameters $r$ and $R$ \citep[see][]{Bowick_PRE_2004,Giomi_EPJE_2008,Jesenek_SM_2015}. We will further elaborate on the relation to the experiments in \citet{Ellis_NP_2018} below. 

\begin{figure}
    \begin{minipage}[t]{.033\textwidth}
        (a)
    \end{minipage}%
    \begin{minipage}[t]{.35\textwidth}
        \centering
        \vspace{0pt}
        \adjincludegraphics[valign=t,width=\textwidth]{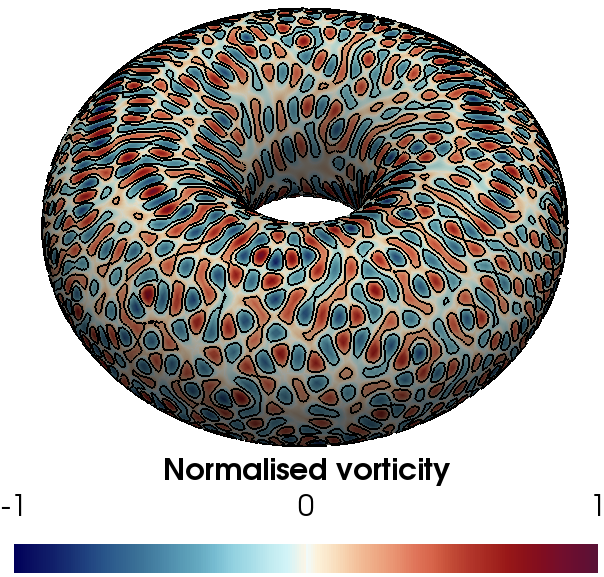}
    \end{minipage}
    \hfill
    \begin{minipage}[t]{.033\textwidth}
        (b)
    \end{minipage}%
    \begin{minipage}[t]{.5\textwidth}
        \centering
        \vspace{0pt}
        \adjincludegraphics[valign=t,width=\textwidth]{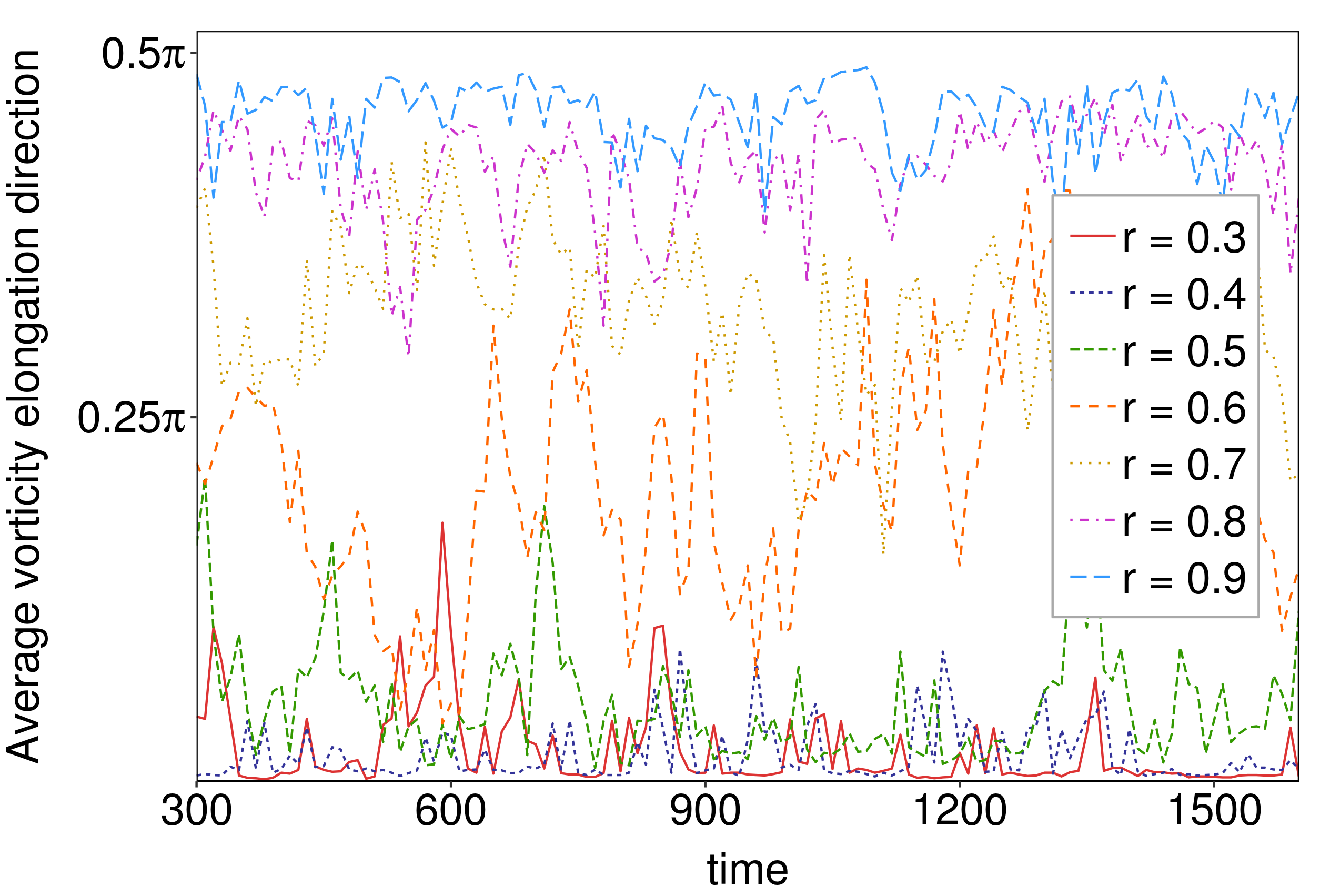}
    \end{minipage} \\~\\
    \begin{minipage}[t]{.033\textwidth}
        (c)
    \end{minipage}%
    \begin{minipage}[t]{.35\textwidth}
        \centering
        \vspace{0pt}
        \adjincludegraphics[valign=t,width=\textwidth]{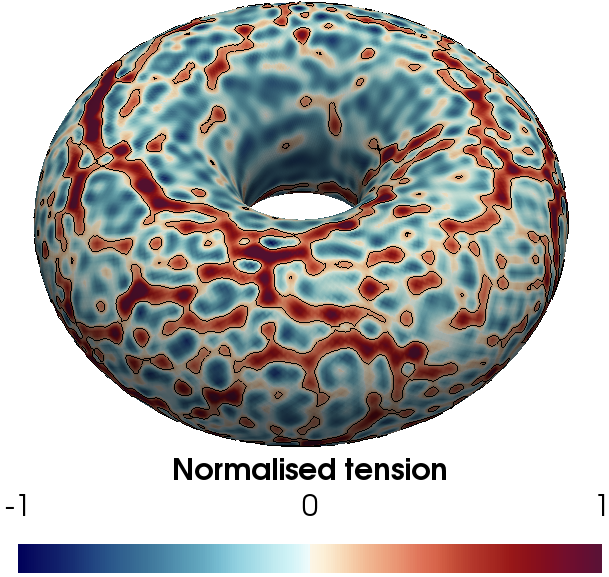}
    \end{minipage}
    \hfill
    \begin{minipage}[t]{.033\textwidth}
        (d)
    \end{minipage}%
    \begin{minipage}[t]{.5\textwidth}
        \centering
        \vspace{0pt}
        \adjincludegraphics[valign=t,width=\textwidth]{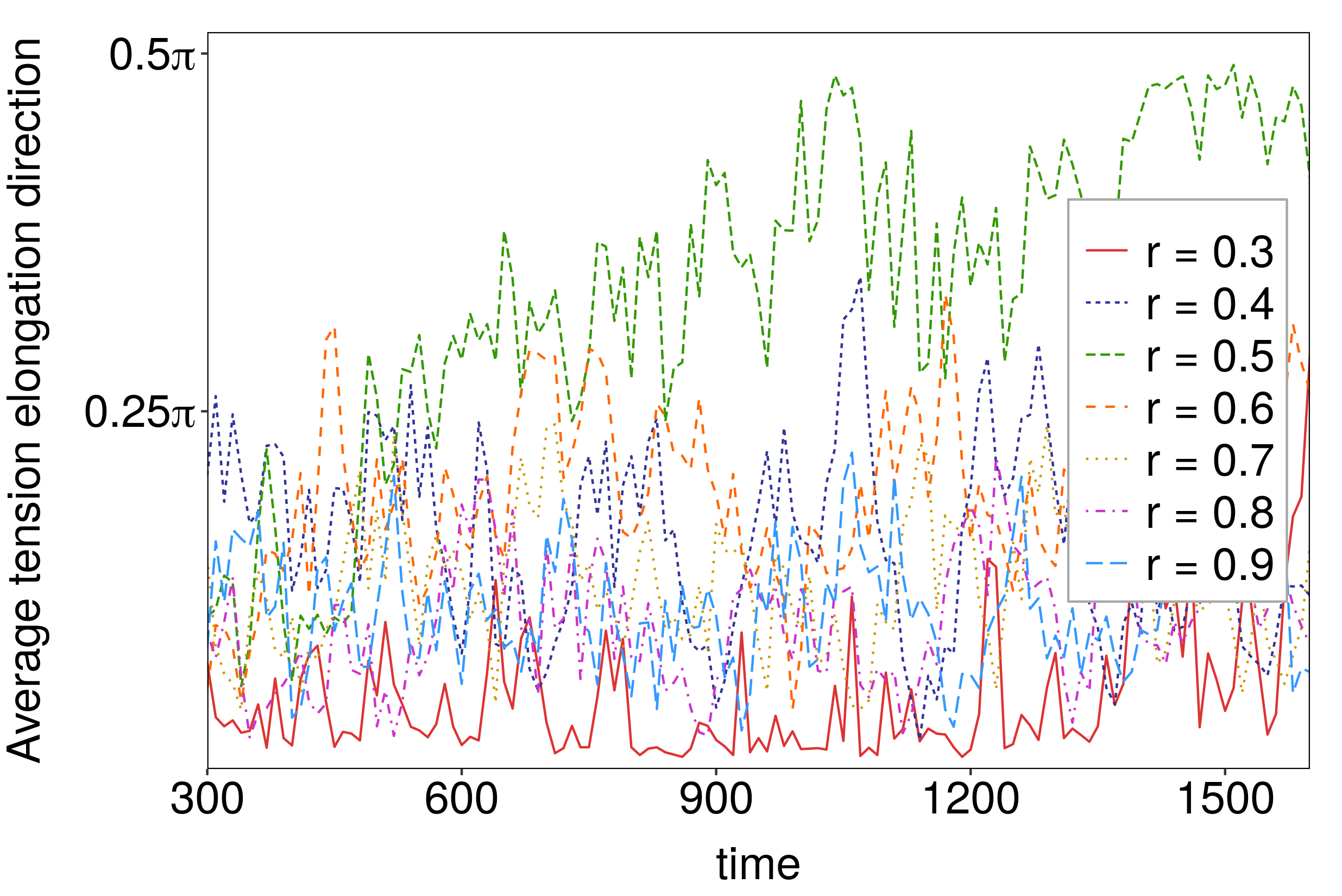}
    \end{minipage}
    \caption{Average direction of elongation of chained vorticity (b) and branched high-tension structures (b) on different tori over time. The elongation direction is considered with respect to the parameterisation. Snapshots of the considered vorticity and surface tension structures for a torus with inner radius $r = 0.8$ are shown in (a) and (c), respectively. For still images of the other tori see Appendix.}
    \label{fig:elongation}
\end{figure}

We first address an observation in the movies in the Electronic Supplement. They indicate a preferred direction of the chained vortex structures. They seem to align with the principle curvature lines and prefer the one with lower absolute value of Gaussian curvature. For regions of positive Gaussian curvature (outer part) they align horizontally, whereas for negative Gaussian curvature (inner part), at least for strong values, the alignment is vertically. This observation is confirmed in Figure \ref{fig:elongation}(b) showing the average direction of elongation of the chained vortex structures. The still image in Figure \ref{fig:elongation}(a) can only partly confirm this, we therefore refer to the corresponding movies in the Electronic Supplement. Figure \ref{fig:elongation}(d) and (c), show the corresponding average direction of elongation of high-tension domains and a characteristic still image, respectively. Representative still images for the other radii are provided in the Appendix. The average directions have been computed using a method first introduced for quantifying deformations of foam structures by \citet{asipauskas} and was later applied for the elongation of cellular structures \citep{activecells}. To apply it for our situation we represent the geometry of the chained vortex structures and the branched high-tension field as phase fields, i.e. scalar fields that take values one on the inside and zero on the outside of the segmented structures with a smooth transition over a small interface. All information about the elongation can be represented in terms of the gradient of these phase fields. The elongation is described as the angle against the lines of constant Gaussian curvature, i.e. the green lines in Figure \ref{fig:torus parametrization}. Thus, an elongation of zero represents a horizontal structure aligned with the green lines and a values of $\pi/2$ represents a vertical structure aligned with the red lines of Figure \ref{fig:torus parametrization}. As we solely want to determine the elongation of large structures, we only consider areas that are bigger than the 99\% quantile of all areas. At least for thin tori ($r=0.3,0.4$) and thick tori ($r=0.8,0.9$) this preferred alignment of the chained vortex structures becomes evident. The elongation of the tension fields is less evident. Only for thin tori ($r=0.3, 0.4$) a preferred horizontal elongation is observed. A preferred vertical elongation for thick tori ($r=0.8, 0.9$) can not be seen. This might result from less dominant branched structures in regions of strong negative Gaussian curvature, see Figure \ref{fig: Heatmaps}(b),(d). For moderate tori ($r=0.5,0.6,0.7$) the average elongation direction of the tension fields fluctuates much stronger with no preferred orientation. Similar alignment effects with minimal curvature lines have also been reported for surface liquid crystals \citep{Segattietal_M3AS_2016,Nestler_JNS_2018,Nestler_SM_2020,Pearce_NJP_2020}, at least if extrinsic curvature effects are taken into account in the surface models \citep{Nestler_SM_2020,Pearce_NJP_2020}. 

Besides this alignment effect with minimal curvature lines the deformation of the vortex structures is also affected by the geometry of the tori. For the chained vortex structures the according mean elongation eigenvalues over all structures of the computations are shown in Table \ref{tab: elongation2}. It represents the mean deformation of the structures, lower values indicating more circular patterns. 
The values are largest for more extreme cases, thin tori ($r=0.3, 0.4$) and thick tori ($r=0.8, 0.9$), and lowest for moderate tori ($r=0.5, 0.6, 0.7$).

\begin{table}
    \centering
    \begin{tabular}{lccccccc}
        Inner radius $r$ & 0.3 & 0.4 & 0.5 & 0.6 & 0.7 & 0.8 & 0.9 \\
        Vorticity elongation eigenvalue & 0.086 & 0.081 & 0.066 & 0.053 & 0.051 & 0.068 & 0.083 \\
    \end{tabular}
    \caption{Elongation eigenvalue averaged over time and simulation runs for tori with inner radius $r$.}
    \label{tab: elongation2}
\end{table}

\begin{figure}
    \begin{minipage}[t]{.033\textwidth}
        (a)
    \end{minipage}%
    \begin{minipage}[t]{.45\textwidth}
        \centering
        \vspace{0pt}
        \adjincludegraphics[valign=t,width=\textwidth]{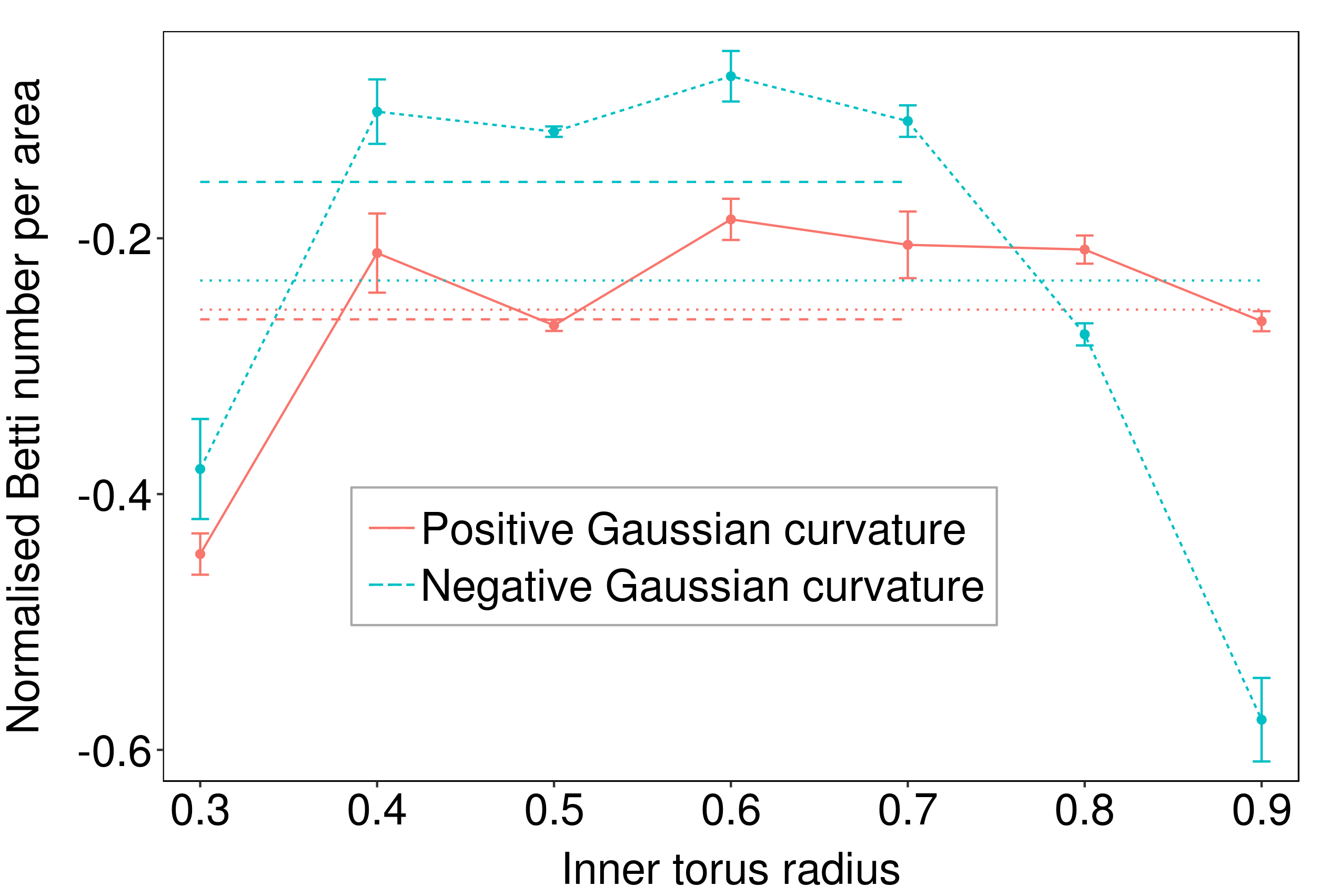}
    \end{minipage}
    \hfill
    \begin{minipage}[t]{.033\textwidth}
        (b)
    \end{minipage}%
    \begin{minipage}[t]{.45\textwidth}
        \centering
        \vspace{0pt}
        \adjincludegraphics[valign=t,width=\textwidth]{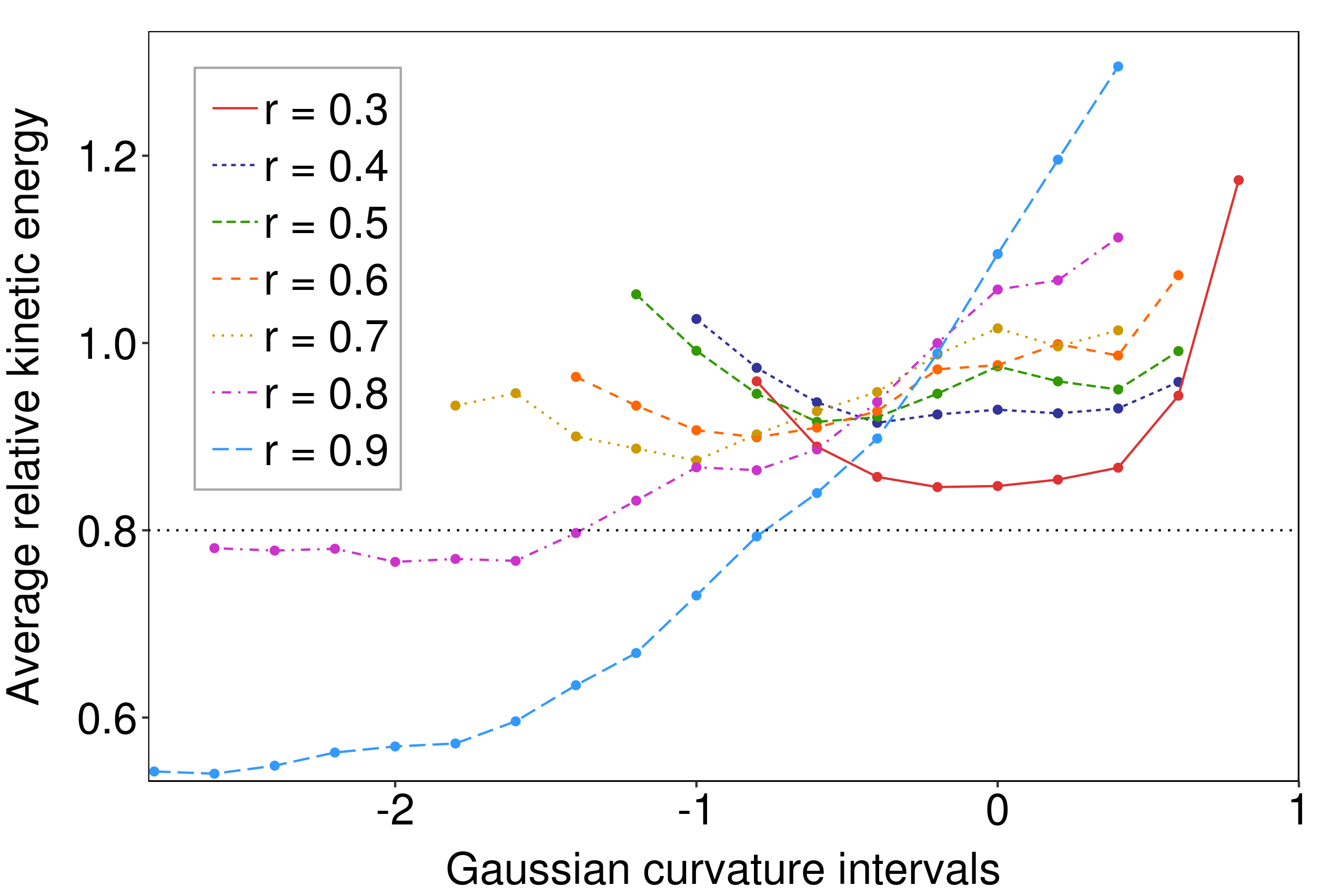}
    \end{minipage}
    \caption{(a) Normalised Betti numbers per area vs. inner torus radius for surface areas with positive and negative Gaussian curvature obtained with parameter $\alpha_\phi = 0.5$. Dotted lines show the means over all tori and dashed lines show the according means over all tori with $r\in\{0.3,0.4,0.5,0.6,0.7\}$. (b) Average total kinetic energy for surface areas with Gaussian curvature $(\kappa-0.1,\kappa+0.1)$, $\kappa \in \{-2.8,-2.6,...,0.8\}$ divided by overall mean total kinetic energy of each torus.}
    \label{fig: Betti positive negative}
\end{figure}

While the previous results mainly indicate a general dependency on the absolute value of the Gaussian curvature, there are also indications on a dependency on the sign of the Gaussian curvature. The only available experimental results for active systems on toroidal surfaces \citep{Ellis_NP_2018} clearly show a dependency on the sign. The corresponding simulations \citep{Pearce_PRL_2019}, with the limitations discussed in the introduction, qualitatively show a higher density of
vortices in the interior of the torus. In the considered active nematic system this is quantified by the topological charge density, defect creation and annihilation rates, which are shown to depend approximately linearly on Gaussian curvature. The GNS equation is a minimal phenomenological model, unable to resolve these details. However, we can measure the number of vortices per area. Figure \ref{fig: Betti positive negative}(a) shows the normalised Betti number per area for positive and negative Gaussian curvature regions for all considered tori. With the exception for thick tori ($r = 0.8, 0.9$) we observe the same behaviour, the number of vortices in the interior of the tori is higher. This confirms a dependency also on the sign of the Gaussian curvature. The discrepancy for thick tori ($r = 0.8, 0.9$) results from the different magnitudes of the vorticity regions that are counted by the Betti number. Figure \ref{fig: Betti positive negative}(b) shows the average total kinetic energy for different regions of the tori. We can observe significantly lower values for regions of strong negative Gaussian curvature. This results in lower normalised Betti numbers for thick tori ($r=0.8,0.9$), as more vorticies fall below the considered threshold. Figure \ref{fig: Betti positive negative}(a) also shows the average values over all tori, with and without the two thick tori ($r=0.8,0.9$). Lower kinetic energy for regions with negative Gaussian curvature and higher values for regions with positive Gaussian curvature can already be seen for the surface NS equation in Figure \ref{fig:energy diagram torus non active}. The effect probably results form the $+ \Gamma_0 \kappa \mathbf{u}$ term in eq. \eqref{eq:active ns}, which, if neglecting all other contributions, leads to exponential decay for $\kappa < 0$ and exponential growth for $\kappa > 0$. It has to remain speculative if this simple explanation can also be used for the highly nonlinear surface GNS equation. Another interesting aspect of Figure \ref{fig: Betti positive negative}(b) are the values at $\kappa = 0$. The differences for different tori indicate not only a dependency on the local Gaussian curvature but also on its gradient, which is largest for thin tori $r=0.3$ and decreases with increasing $r$. Again this behaviour is in accordance to the experiments in \citet{Ellis_NP_2018}. 

\section{Conclusions}\label{sec:conclusion}
  
We consider a surface generalised Navier-Stokes (GNS) equation as a minimal model for active flows on arbitrarily curved surfaces. This extends work of \citet{Mickelin_PRL_2018}, who considered this equation on a sphere. The numerical approach extends work of \citet{Reuther_PF_2018} for the surface Navier-Stokes (NS) equation and is based on the general concept to solve surface vector-valued partial differential equations on arbitrary surfaces by surface finite elements \citep{Nestler_JCP_2019}. We focus on toroidal surfaces, as a prototypical example of surfaces with varying Gaussian curvature with positive and negative values and consider parameter settings which lead to anomalous chained turbulence on a sphere \citep{Mickelin_PRL_2018}. We here concentrate on the influence of Gaussian curvature on this new regime of active turbulence. The simulation results suggest that this turbulence regime can be influenced by global properties of the surface but also via the local Gaussian curvature and its gradients. The chained vortex structures have the tendency to align with minimal curvature lines of the surface. At the outer part (positive Gaussian curvature) they show a tendency to align horizontally and at the inner part (negative Gaussian curvature) they seem to align vertically, at least for thick tori. The considered topological and geometrical measures for the vorticity and high-tension fields, the normalised Betti and Branch numbers, indicate anomalous turbulence for moderate values of Gaussian curvature and possible deviations for lower and higher values. Also the kinetic energy and the enstrophy not only depend on the global properties of the torus but also on the local Gaussian curvature. Their ratio, another, at least qualitative, measure for anomalous turbulence \citep{Mickelin_PRL_2018} only slightly deviates for more extreme values for the inner radius $r$ or the local Gaussian curvature $\kappa$, and shows the same weak dependency on curvature as the considered topological and geometric measures.

While a full understanding of the relation of Gaussian curvature on active turbulence requires many further investigations, the simulations results indicate a clear dependency of various aspects on the Gaussian curvature of the surface. Some are in qualitative accordance with the experiments on active nematic liquid crystals, which are constrained to lie on a toroidal surface, see \citet{Ellis_NP_2018}. These are larger numbers of vortices in regions of negative Gaussian curvature and a dependency on the gradient in Gaussian curvature on the kinetic energy. Other effects, such as the alignment of the chained structures with minimal curvature lines, ask for experimental validation. Another interesting question for future research is the influence of local Gaussian curvature on the transitions to classical 2D Kolmogorov turbulence. This transition is addressed in flat space and on the sphere  \citep[see][]{Bratanovetal_PNAS_2015,Jamesetal_PRF_2018,Mickelin_PRL_2018,Linkmannetal_PRL_2019,Linkmannetal_PRE_2020,Linkmannetal_JFM_2020}. \\

Acknowledgment: This research was supported by the German Research Foundation (DFG) within the Research Unit 3013. We used computing resources provided by ZIH at TU Dresden and by J\"ulich Supercomputing Centre within project HDR06. We further acknowledge the provided data from \citet{Mickelin_PRL_2018} by O. Mickelin and J. Dunkel, as well as support from M. Nestler, M. Salvalaglio and D. Wenzel concerning the postprocessing.

\appendix
\section{}\label{appendix}

\begin{figure}
    \begin{minipage}[t]{.033\textwidth}
        (a)
    \end{minipage}%
    \begin{minipage}[t]{.44\textwidth}
        $\alpha_\phi=0.4$
        \centering
        \vspace{0pt}
        \adjincludegraphics[valign=t,width=\textwidth]{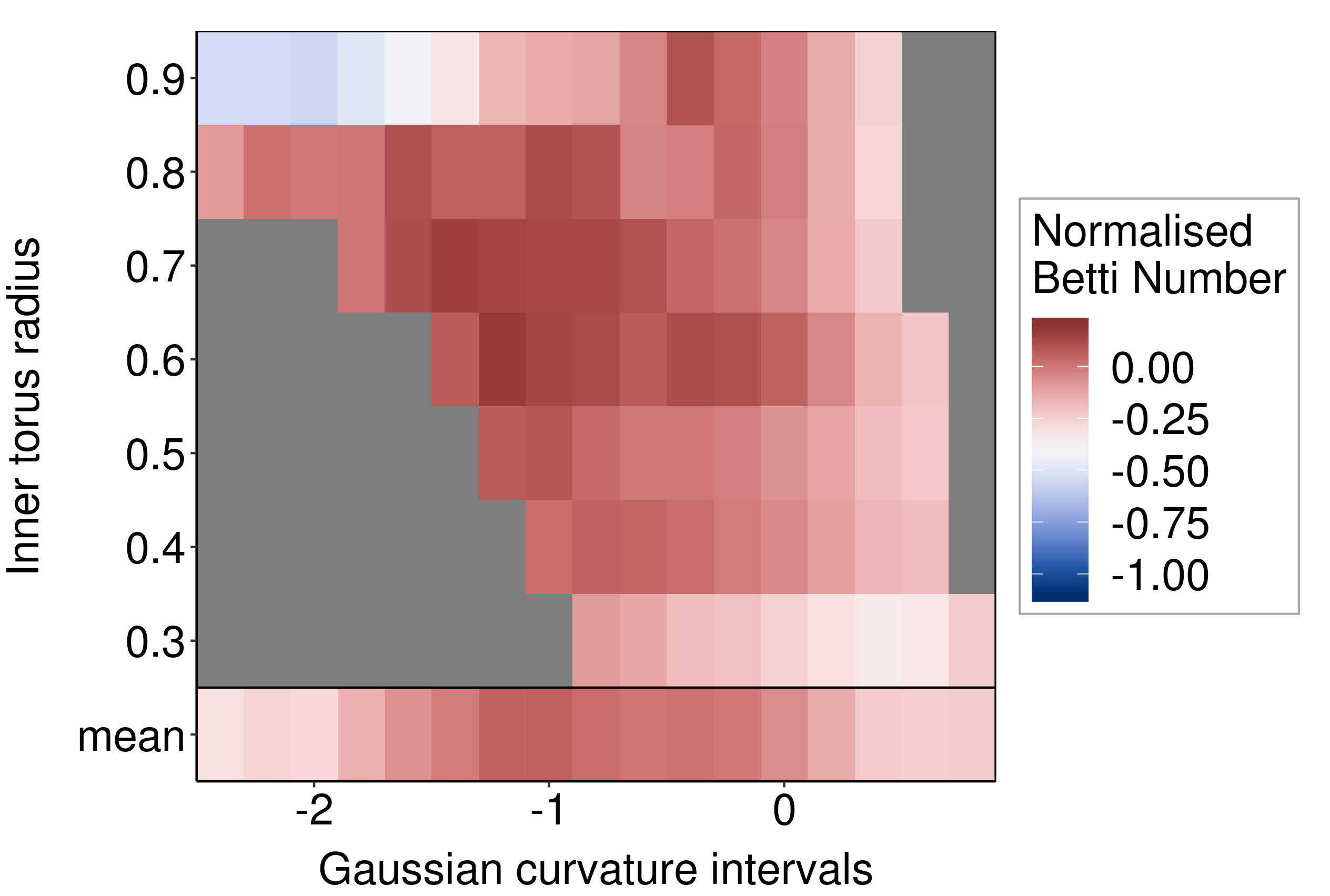}
    \end{minipage}%
    \hfill
    \begin{minipage}[t]{.033\textwidth}
        (b)
    \end{minipage}%
    \begin{minipage}[t]{.44\textwidth}
        $\alpha_\phi=0.5$
        \centering
        \vspace{0pt}
        \adjincludegraphics[valign=t,width=\textwidth]{results/torus_active/betti_heatmap_R_0.5.png}
    \end{minipage} \\
    \begin{minipage}[t]{.033\textwidth}
        (c)
    \end{minipage}%
    \begin{minipage}[t]{.44\textwidth}
        $\alpha_\phi=0.6$
        \centering
        \vspace{0pt}
        \adjincludegraphics[valign=t,width=\textwidth]{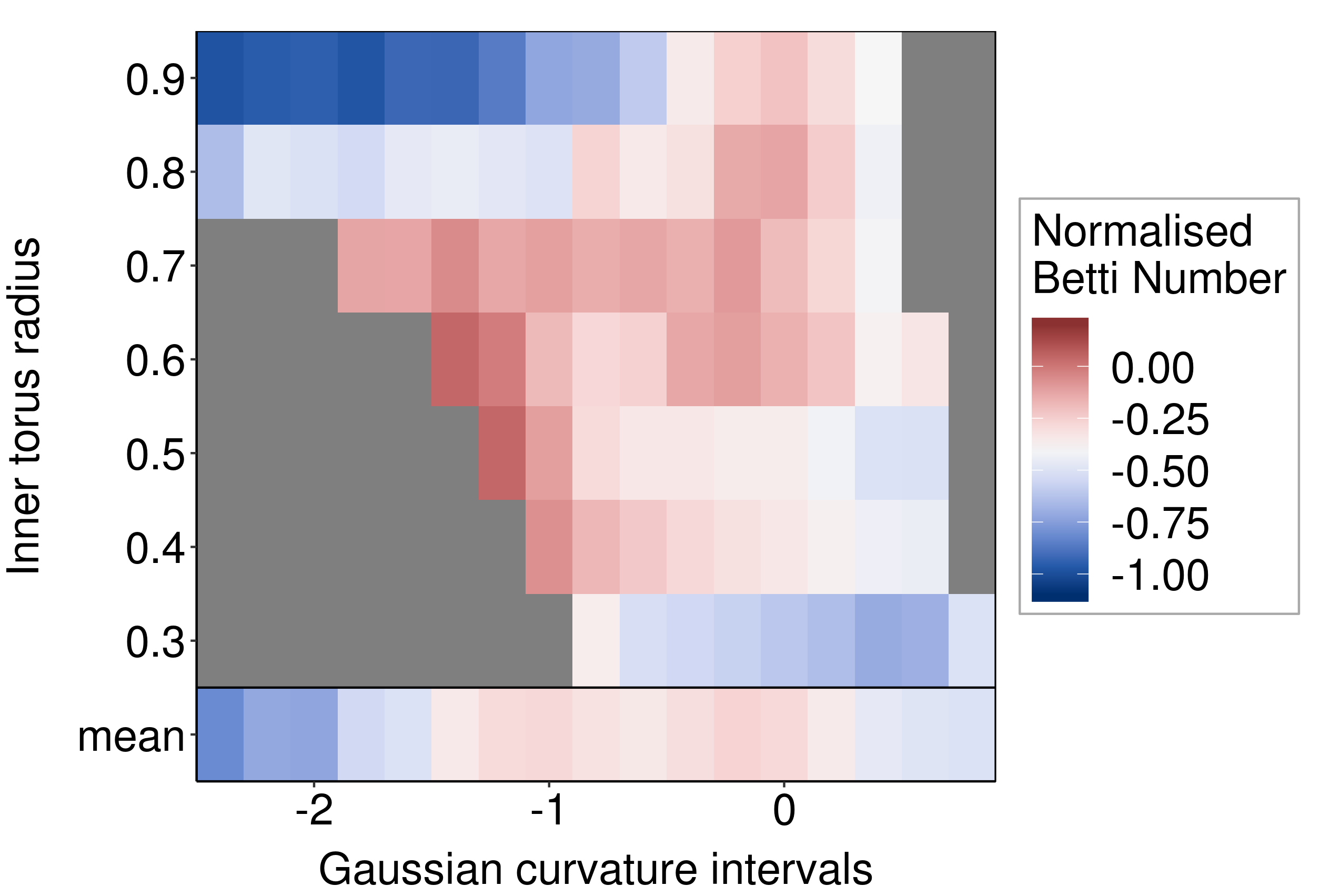}
    \end{minipage}
    \hspace*{\fill}
    \begin{minipage}[t]{.033\textwidth}
        (d)
    \end{minipage}%
    \begin{minipage}[t]{.44\textwidth}
        $\alpha_\phi=0.75$
        \centering
        \vspace{0pt}
        \adjincludegraphics[valign=t,width=\textwidth]{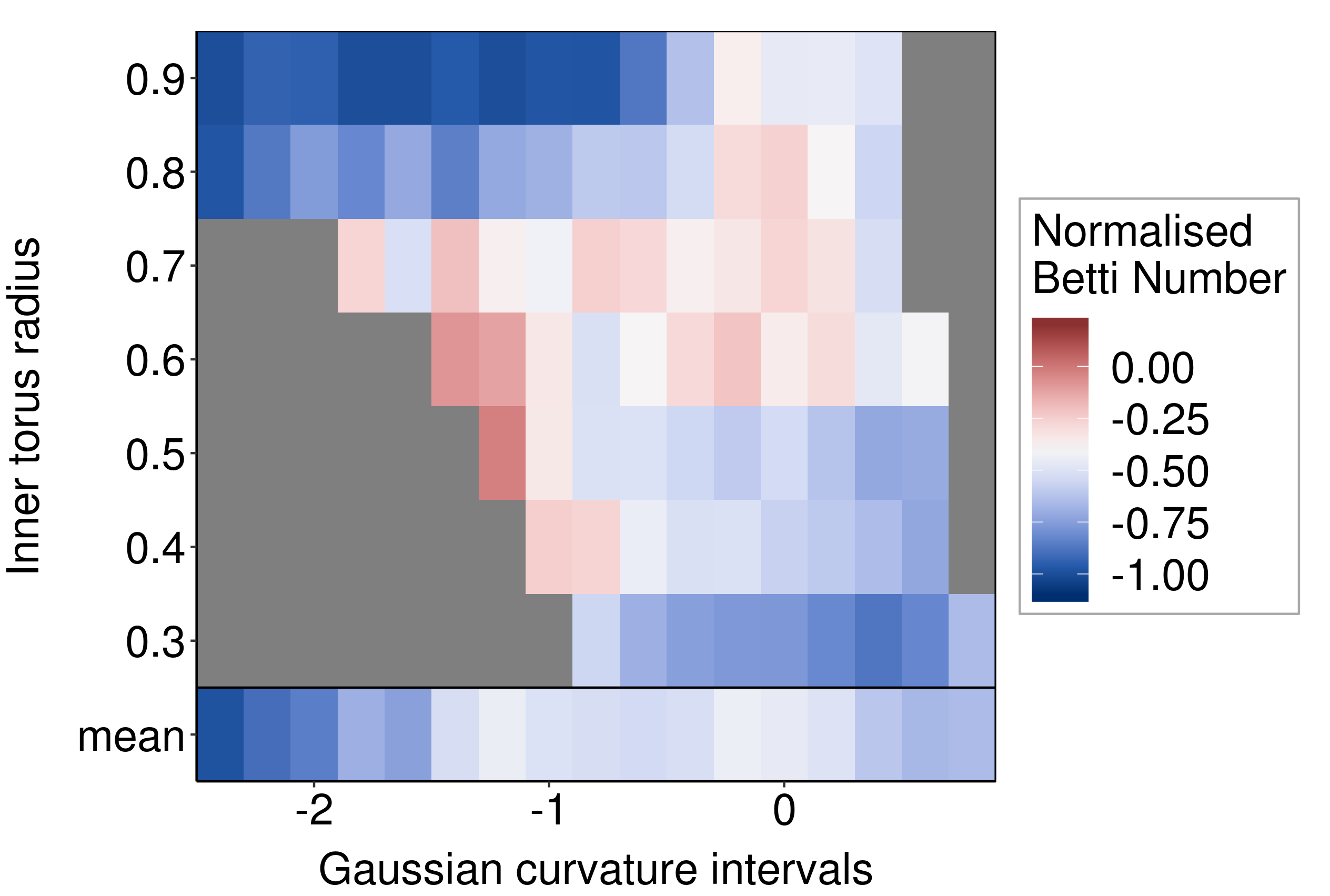}
    \end{minipage}
    \caption{Normalised Betti numbers per area as in Figure \ref{fig: Heatmaps}(a) for different thresholds to define regions with high absolute vorticity $\{\x\in M:\phi(\x,t) > \alpha_\phi\cdot\max_{\x\in M}\phi(\x,t) \textrm{ or } \phi(\x,t) < \alpha_\phi\cdot\min_{\x\in M}\phi(\x,t)\}$, with (a) $\alpha_\phi=0.4$, (b) $\alpha_\phi=0.5$, (c) $\alpha_\phi=0.6$ and (d) $\alpha_\phi=0.75$.}
    \label{fig: Betti different thresholds}
\end{figure}

\begin{figure}
    \begin{minipage}[t]{.033\textwidth}
        (a)
    \end{minipage}%
    \begin{minipage}[t]{.44\textwidth}
        $\beta_p=0.8$
        \centering
        \vspace{0pt}
        \adjincludegraphics[valign=t,width=\textwidth]{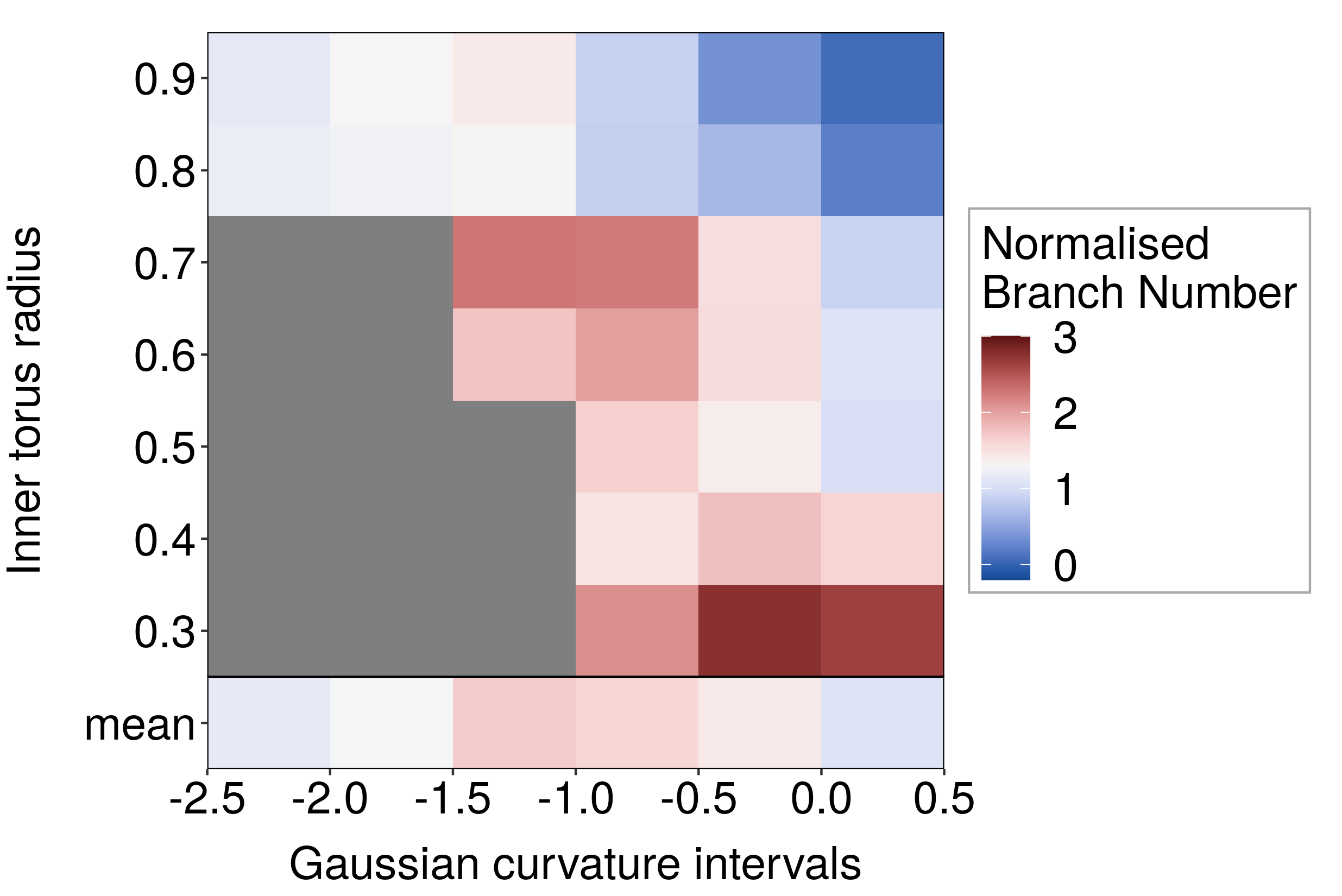}
    \end{minipage}%
    \hfill
    \begin{minipage}[t]{.033\textwidth}
        (b)
    \end{minipage}%
    \begin{minipage}[t]{.44\textwidth}
        $\beta_p=0.9$
        \centering
        \vspace{0pt}
        \adjincludegraphics[valign=t,width=\textwidth]{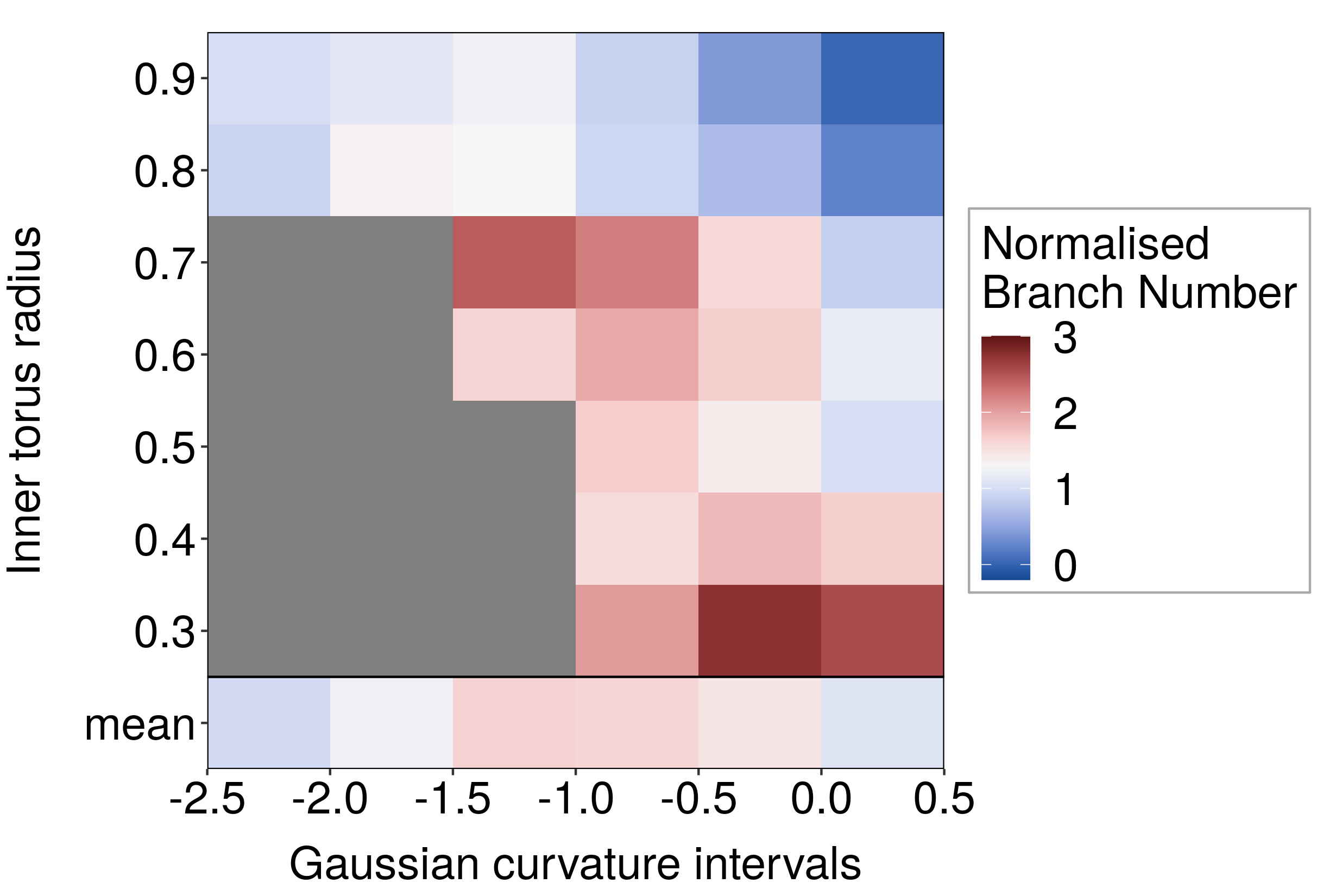}
    \end{minipage} \\
    \begin{minipage}[t]{.033\textwidth}
        (c)
    \end{minipage}%
    \begin{minipage}[t]{.44\textwidth}
        $\beta_p=1.1$
        \centering
        \vspace{0pt}
        \adjincludegraphics[valign=t,width=\textwidth]{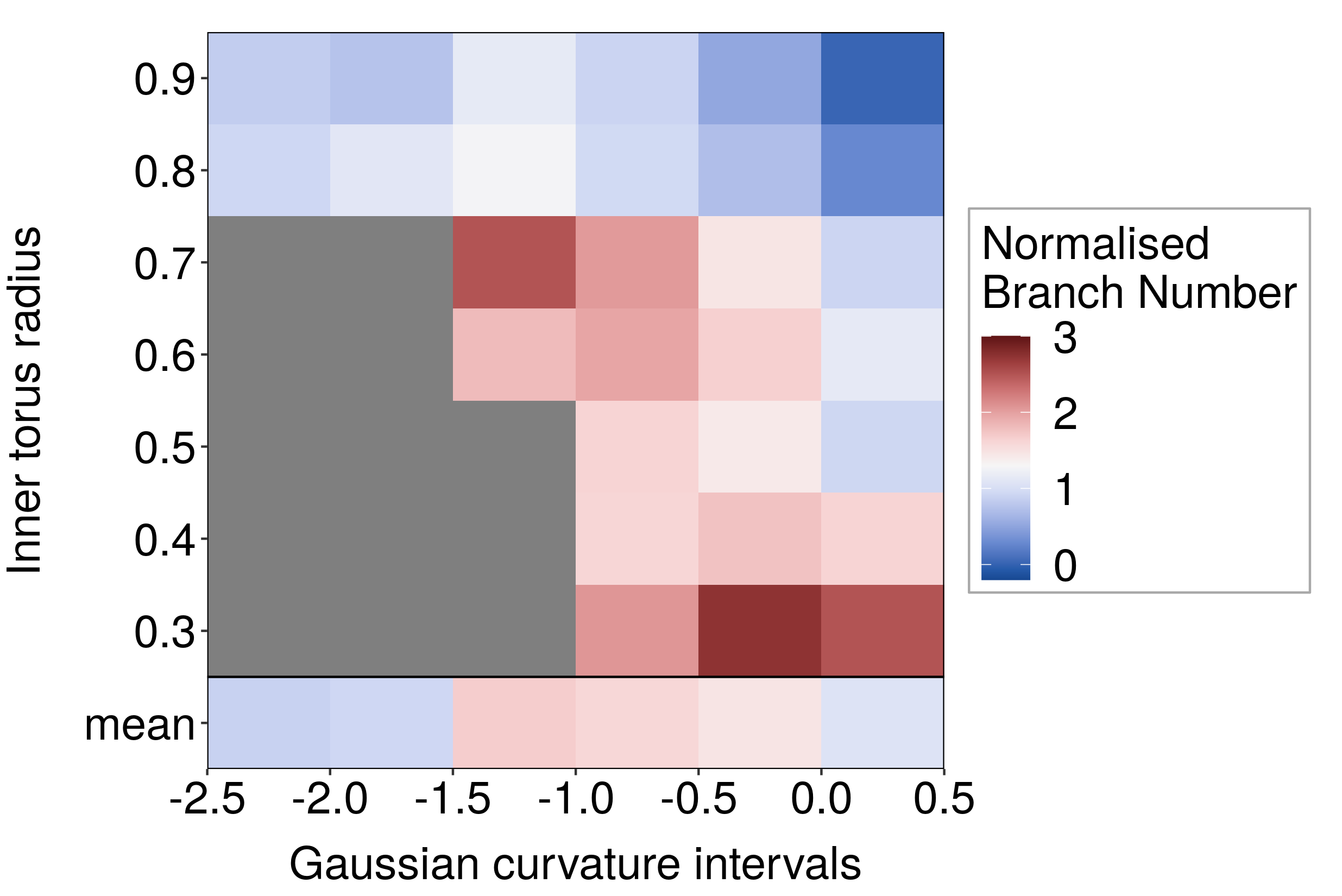}
    \end{minipage}
    \hspace*{\fill}
    \begin{minipage}[t]{.033\textwidth}
        (d)
    \end{minipage}%
    \begin{minipage}[t]{.44\textwidth}
        $\beta_p=1.2$
        \centering
        \vspace{0pt}
        \adjincludegraphics[valign=t,width=\textwidth]{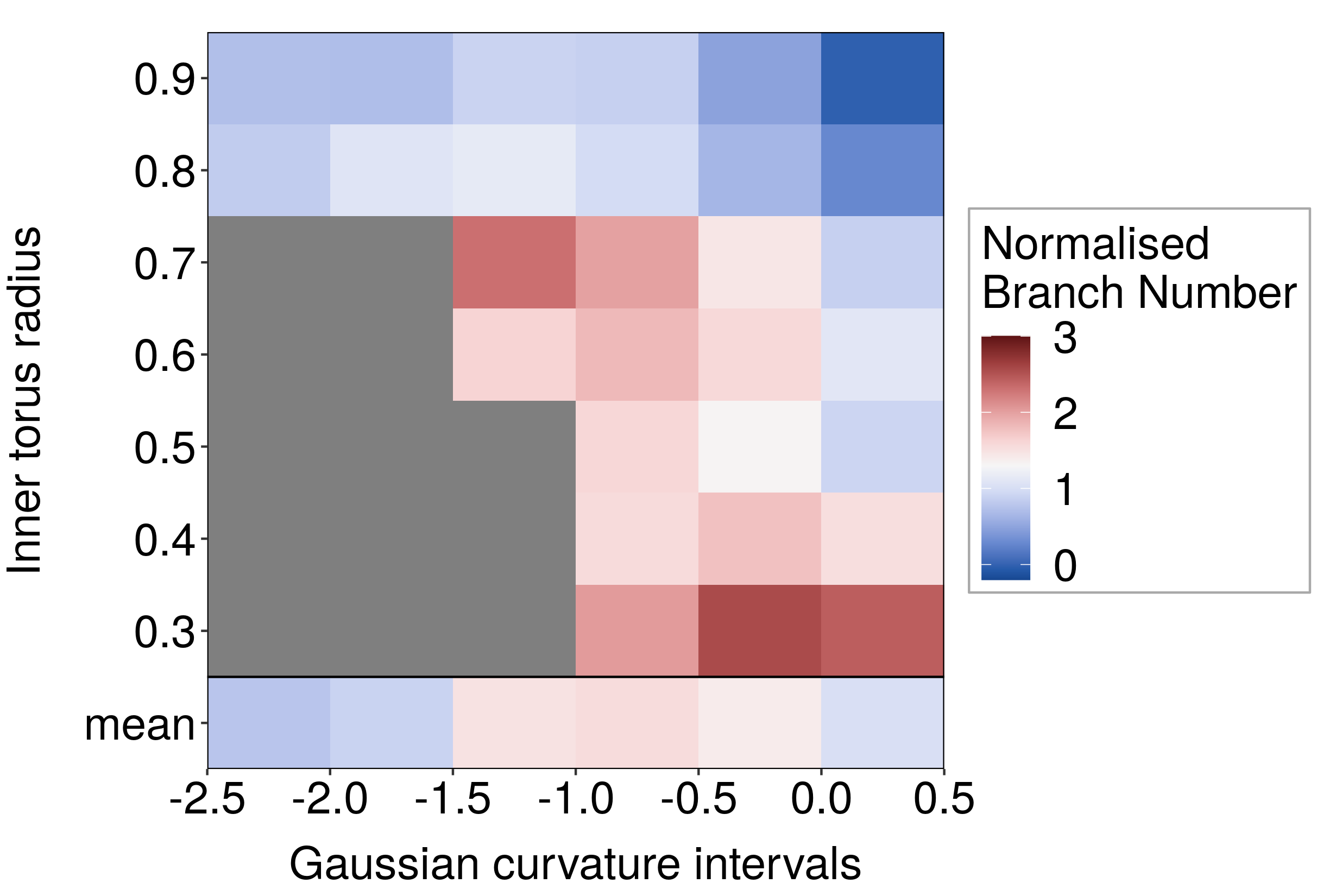}
    \end{minipage}
    \caption{Normalised Branch numbers per area as in Figure \ref{fig: Heatmaps}(b) for different regions $\{x\in M:p(\x,t) > \beta_p \cdot\Bar{p}(t)\}$ with (a) $\beta_p=0.8$, (b) $\beta_p=0.9$, (c) $\beta_p=1.1$ and (d) $\beta_pa=1.2$.}
    \label{fig: Branch different thresholds}
\end{figure}

\begin{figure}
    \begin{minipage}[t]{.032\textwidth}
        (a)
    \end{minipage}%
    \begin{minipage}[t]{.21\textwidth}
        $r = 0.3$
        \centering
        \vspace{0pt}
        \adjincludegraphics[valign=t,width=\textwidth]{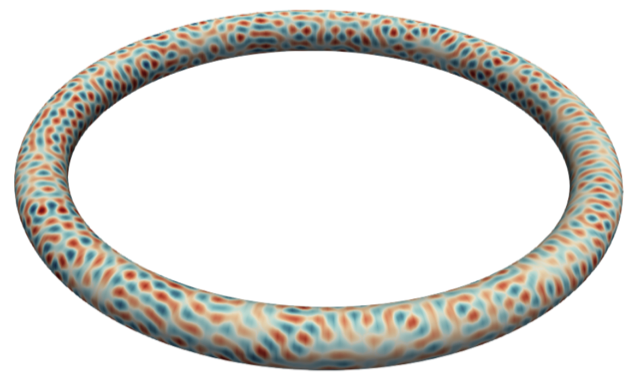}
    \end{minipage}%
    \hfill
    \begin{minipage}[t]{.032\textwidth}
        (b)
    \end{minipage}%
    \begin{minipage}[t]{.21\textwidth}
        $r = 0.4$
        \centering
        \vspace{0pt}
        \adjincludegraphics[valign=t,width=\textwidth]{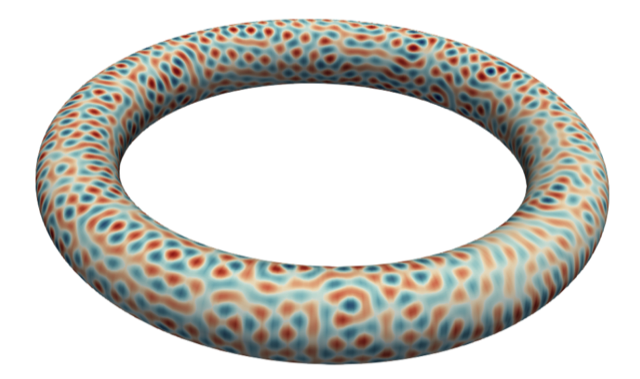}
    \end{minipage}
    \hfill
    \begin{minipage}[t]{.032\textwidth}
        (c)
    \end{minipage}%
    \begin{minipage}[t]{.21\textwidth}
        $r = 0.5$
        \centering
        \vspace{0pt}
        \adjincludegraphics[valign=t,width=\textwidth]{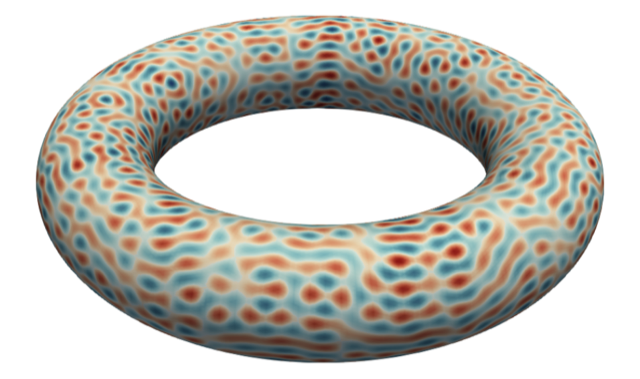}
    \end{minipage}%
    \hfill
    \begin{minipage}[t]{.032\textwidth}
        (d)
    \end{minipage}%
    \begin{minipage}[t]{.21\textwidth}
        $r = 0.6$
        \centering
        \vspace{0pt}
        \adjincludegraphics[valign=t,width=\textwidth]{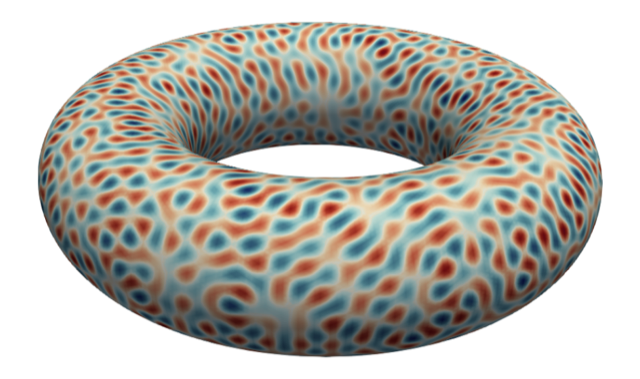}
    \end{minipage}\\~\\
    \begin{minipage}[t]{.032\textwidth}
        (e)
    \end{minipage}%
    \begin{minipage}[t]{.21\textwidth}
        $r = 0.7$
        \centering
        \vspace{0pt}
        \adjincludegraphics[valign=t,width=\textwidth]{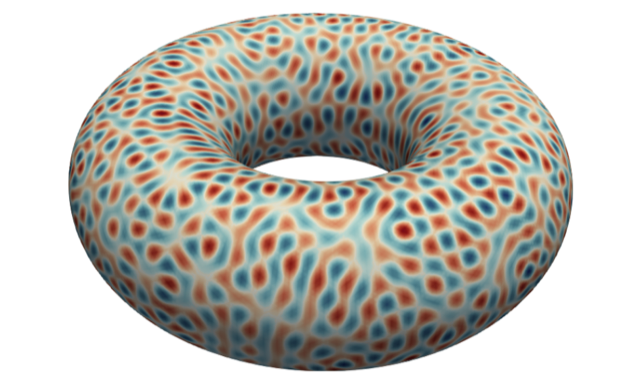}
    \end{minipage}%
    \hfill
    \begin{minipage}[t]{.032\textwidth}
        (f)
    \end{minipage}%
    \begin{minipage}[t]{.21\textwidth}
        $r = 0.8$
        \centering
        \vspace{0pt}
        \adjincludegraphics[valign=t,width=\textwidth]{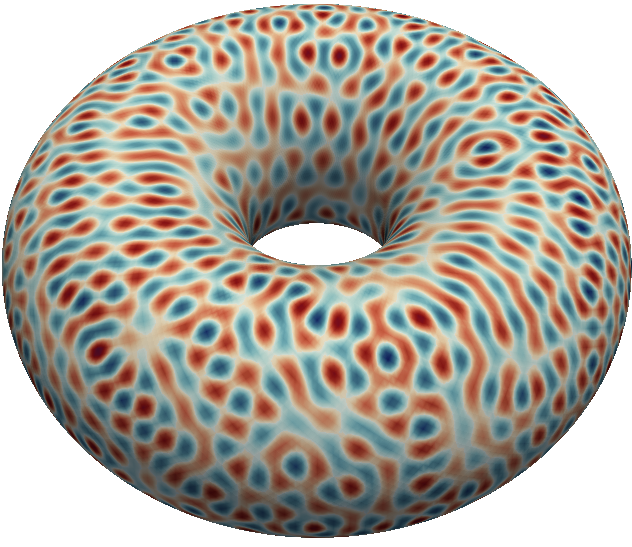}
    \end{minipage}
    \hfill
    \begin{minipage}[t]{.032\textwidth}
        (g)
    \end{minipage}%
    \begin{minipage}[t]{.21\textwidth}
        $r = 0.9$
        \centering
        \vspace{0pt}
        \adjincludegraphics[valign=t,width=\textwidth]{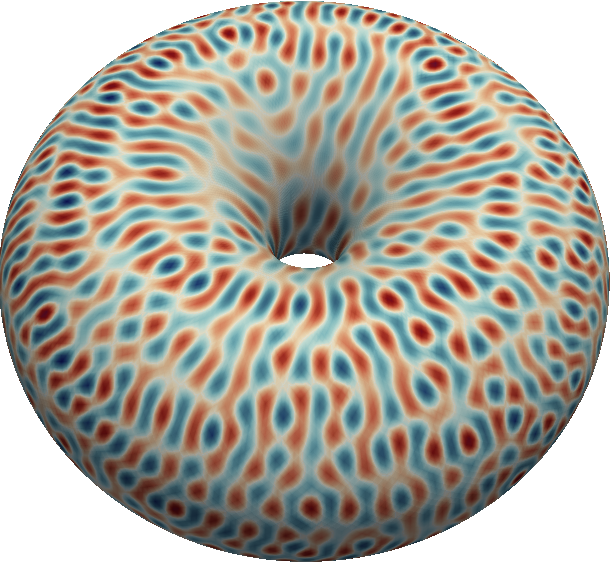}
    \end{minipage}%
    \hfill
    \begin{minipage}[t]{.032\textwidth}
        \textrm{}
    \end{minipage}%
    \begin{minipage}[t]{.21\textwidth}
        \centering
        \vspace{20pt}
        \adjincludegraphics[valign=t,width=\textwidth]{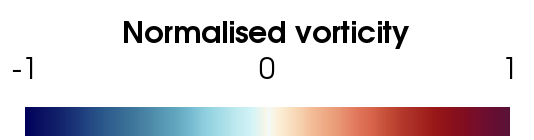}
    \end{minipage}%
    \hfill
    \caption{Snapshots of normalised vorticity fields on different torus surfaces in the active regime.}
    \label{fig:still images vorticity}
\end{figure}

\begin{figure}
    \begin{minipage}[t]{.032\textwidth}
        (a)
    \end{minipage}%
    \begin{minipage}[t]{.21\textwidth}
        $r = 0.3$
        \centering
        \vspace{0pt}
        \adjincludegraphics[valign=t,width=\textwidth]{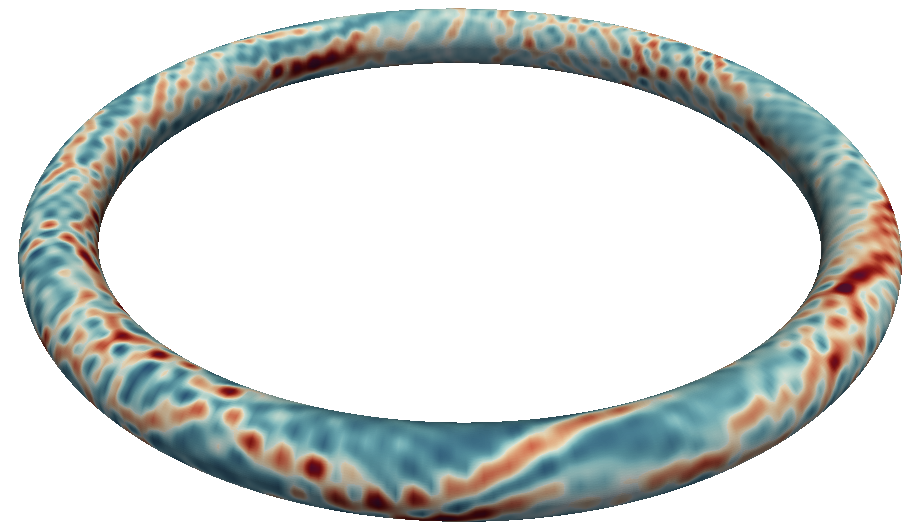}
    \end{minipage}%
    \hfill
    \begin{minipage}[t]{.032\textwidth}
        (b)
    \end{minipage}%
    \begin{minipage}[t]{.21\textwidth}
        $r = 0.4$
        \centering
        \vspace{0pt}
        \adjincludegraphics[valign=t,width=\textwidth]{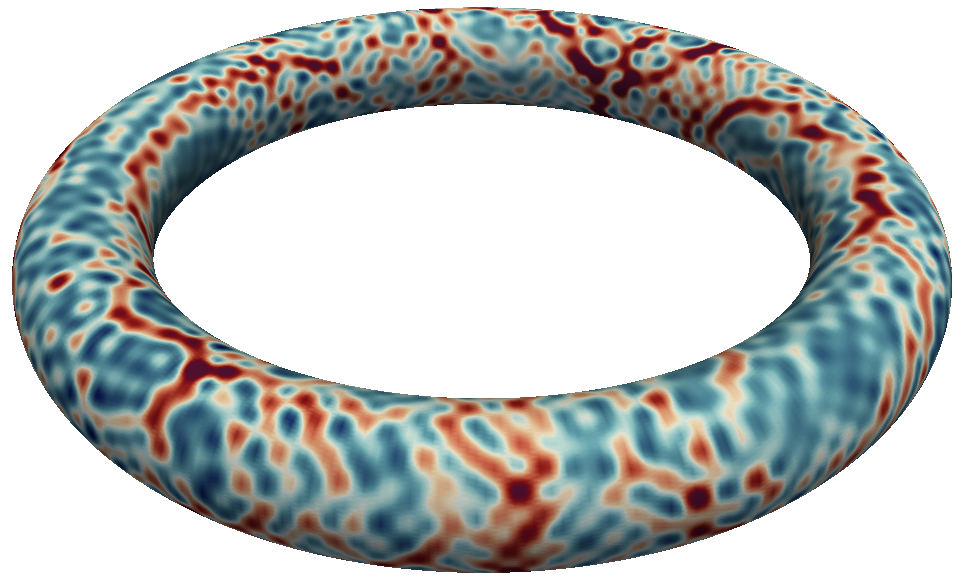}
    \end{minipage}
    \hfill
    \begin{minipage}[t]{.032\textwidth}
        (c)
    \end{minipage}%
    \begin{minipage}[t]{.21\textwidth}
        $r = 0.5$
        \centering
        \vspace{0pt}
        \adjincludegraphics[valign=t,width=\textwidth]{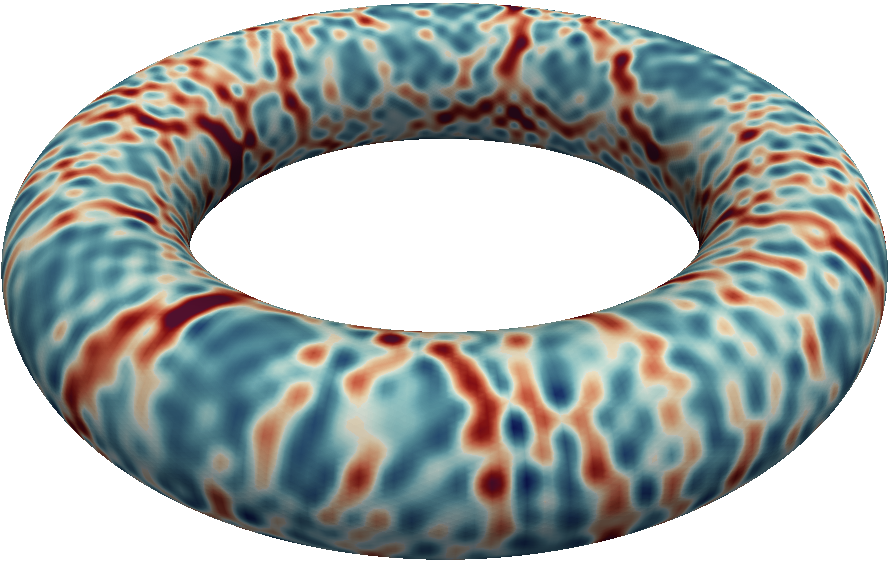}
    \end{minipage}%
    \hfill
    \begin{minipage}[t]{.032\textwidth}
        (d)
    \end{minipage}%
    \begin{minipage}[t]{.21\textwidth}
        $r = 0.6$
        \centering
        \vspace{0pt}
        \adjincludegraphics[valign=t,width=\textwidth]{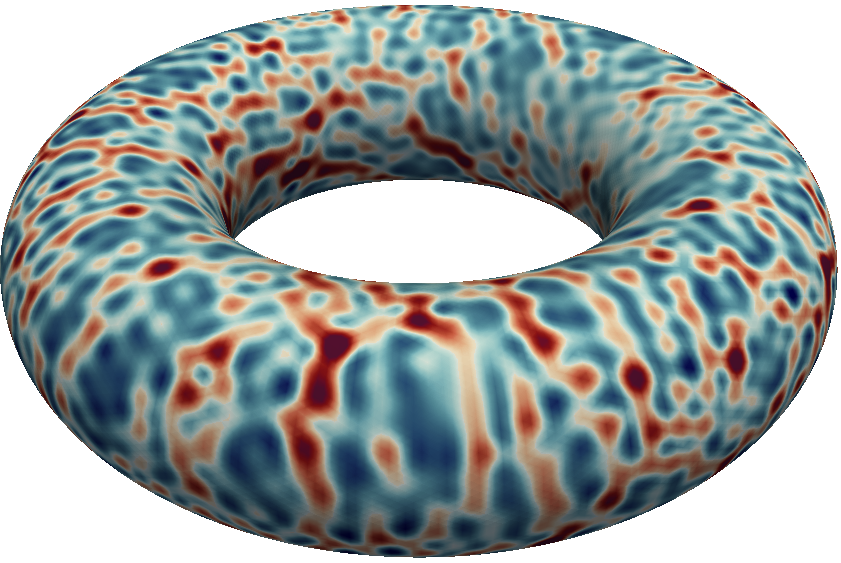}
    \end{minipage}\\~\\
    \begin{minipage}[t]{.032\textwidth}
        (e)
    \end{minipage}%
    \begin{minipage}[t]{.21\textwidth}
        $r = 0.7$
        \centering
        \vspace{0pt}
        \adjincludegraphics[valign=t,width=\textwidth]{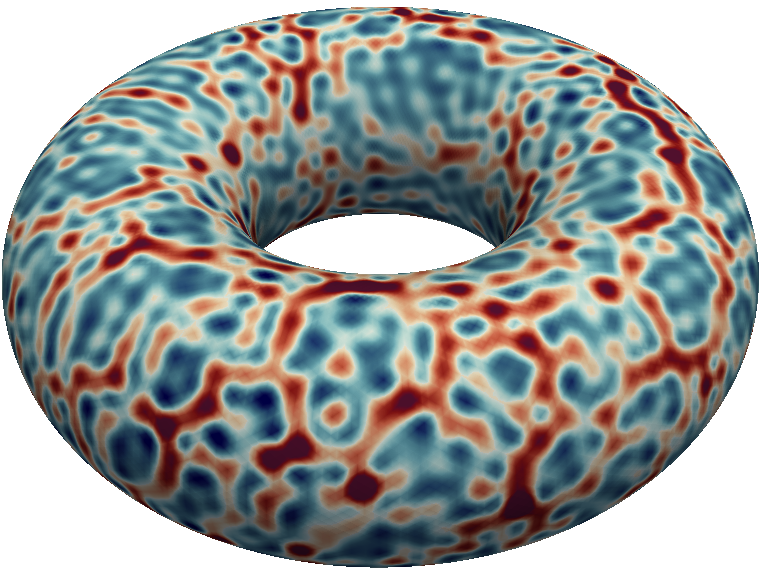}
    \end{minipage}%
    \hfill
    \begin{minipage}[t]{.032\textwidth}
        (f)
    \end{minipage}%
    \begin{minipage}[t]{.21\textwidth}
        $r = 0.8$
        \centering
        \vspace{0pt}
        \adjincludegraphics[valign=t,width=\textwidth]{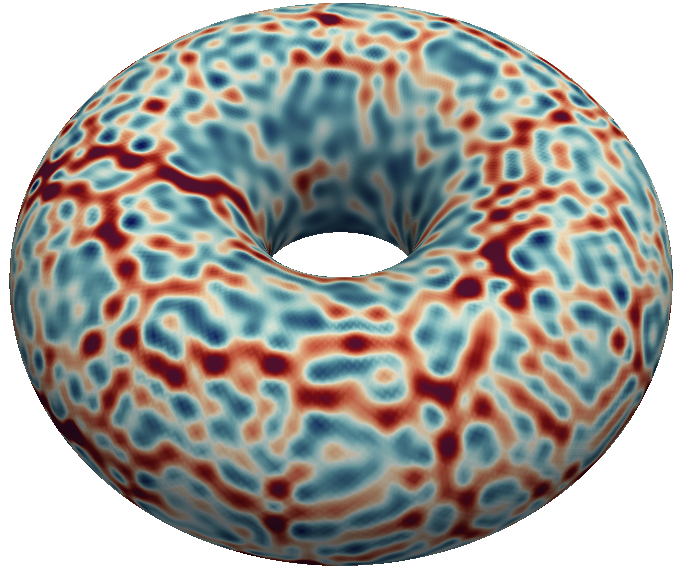}
    \end{minipage}
    \hfill
    \begin{minipage}[t]{.032\textwidth}
        (g)
    \end{minipage}%
    \begin{minipage}[t]{.21\textwidth}
        $r = 0.9$
        \centering
        \vspace{0pt}
        \adjincludegraphics[valign=t,width=\textwidth]{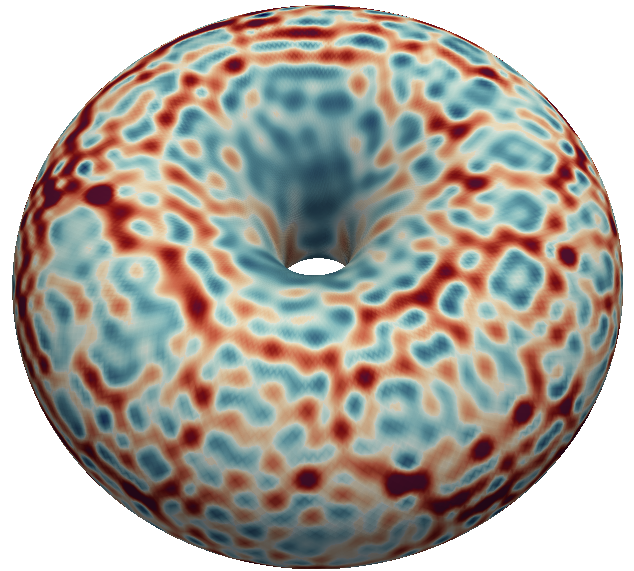}
    \end{minipage}%
    \hfill
    \begin{minipage}[t]{.032\textwidth}
        \textrm{}
    \end{minipage}%
    \begin{minipage}[t]{.21\textwidth}
        \centering
        \vspace{20pt}
        \adjincludegraphics[valign=t,width=\textwidth]{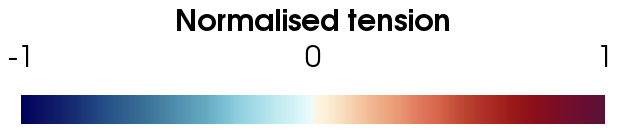}
    \end{minipage}%
    \hfill
    \caption{Snapshots of normalised surface tension fields on different torus surfaces in the active regime.}
    \label{fig:still images tension}
\end{figure}

Figures \ref{fig: Betti different thresholds} and \ref{fig: Branch different thresholds} provide the same information as Figure \ref{fig: Betti positive negative} for different thresholds $\alpha_\phi$ and $\beta_p$, demonstrating the robustness of the results on these values. Figure \ref{fig:still images vorticity} and \ref{fig:still images tension} show the corresponding still images for the vorticity and tension structures in Figure \ref{fig:elongation} for the other tori. For the corresponding videos and additional visualizations of the surface velocity using LIC we refer to the Electronic Supplement.

\bibliographystyle{jfm}
\bibliography{literature}

\end{document}